\documentclass[10pt,a4paper,twoside]{report}

\usepackage[utf8]{inputenc}   

\usepackage[english]{babel} 

\newcommand{\acknowledgments}{@undefined} 
\addto\captionsenglish{\renewcommand{\acknowledgments}{Acknowledgments}}

\addto\captionsportuguese{\renewcommand{\acknowledgments}{Agradecimentos}}
\addto\captionsportuguese{\renewcommand{\nomname}{Lista de S\'{i}mbolos}} 

\def\FontLb{
  \usefont{T1}{phv}{b}{n}\fontsize{16pt}{16pt}\selectfont}

\def\FontMb{
  \usefont{T1}{phv}{b}{n}\fontsize{14pt}{14pt}\selectfont}
\def\FontSn{
  \usefont{T1}{phv}{m}{n}\fontsize{12pt}{12pt}\selectfont}

\usepackage{geometry}	
\usepackage{setspace}

\usepackage{graphicx}

\usepackage{amsmath}  
\usepackage{amsthm}   
\usepackage{amsfonts} %

\usepackage{subfigure}

\usepackage{subfigmat}

\usepackage{dcolumn}
\newcolumntype{d}{D{.}{.}{-1}} 
\newcolumntype{e}{D{E}{E}{-1}} 

\usepackage{nomencl}
\makenomenclature
\RequirePackage{ifthen} 
\ifthenelse{\equal{\languagename}{english}}%
    { 
    \renewcommand{\nomgroup}[1]{%
      \ifthenelse{\equal{#1}{R}}{%
        \item[\textbf{Roman symbols}]}{%
        \ifthenelse{\equal{#1}{G}}{%
          \item[\textbf{Greek symbols}]}{%
          \ifthenelse{\equal{#1}{S}}{%
            \item[\textbf{Subscripts}]}{%
            \ifthenelse{\equal{#1}{T}}{%
              \item[\textbf{Superscripts}]}{}}}}}%
    }{
    \renewcommand{\nomgroup}[1]{%
      \ifthenelse{\equal{#1}{R}}{%
        \item[\textbf{Simbolos romanos}]}{%
        \ifthenelse{\equal{#1}{G}}{%
          \item[\textbf{Simbolos gregos}]}{%
          \ifthenelse{\equal{#1}{S}}{%
            \item[\textbf{Subscritos}]}{%
            \ifthenelse{\equal{#1}{T}}{%
              \item[\textbf{Sobrescritos}]}{}}}}}%
    }%

\usepackage[number=none]{glossary}
\setglossary{gloskip={}}
\makeglossary

\usepackage{rotating}

\usepackage[pdftex]{hyperref} 
\hypersetup{colorlinks,       
            linkcolor=black,  
            anchorcolor=black,
            citecolor=black,  
            filecolor=black,  
            menucolor=black,  
            pagecolor=black,  
            urlcolor=black,   
	          bookmarks=true,         
	          bookmarksopen=false,    
	          bookmarksnumbered=true, 
	          pdftitle={Thesis},
            pdfauthor={Andre C. Marta},
            pdfsubject={Thesis Title},
            pdfkeywords={Thesis Keywords},
            pdfstartview=FitV,
            pdfdisplaydoctitle=true}

\usepackage[figure,table]{hypcap}

\def\be{\begin{equation}}
\def\ee{\end{equation}}
\def\beq{\begin{eqnarray}}
\def\eeq{\end{eqnarray}}

\begin{document}

\pagestyle{plain}

\pagenumbering{roman}


\thispagestyle {empty}

\includegraphics[bb=9.5cm 11cm 0cm 0cm,scale=0.29]{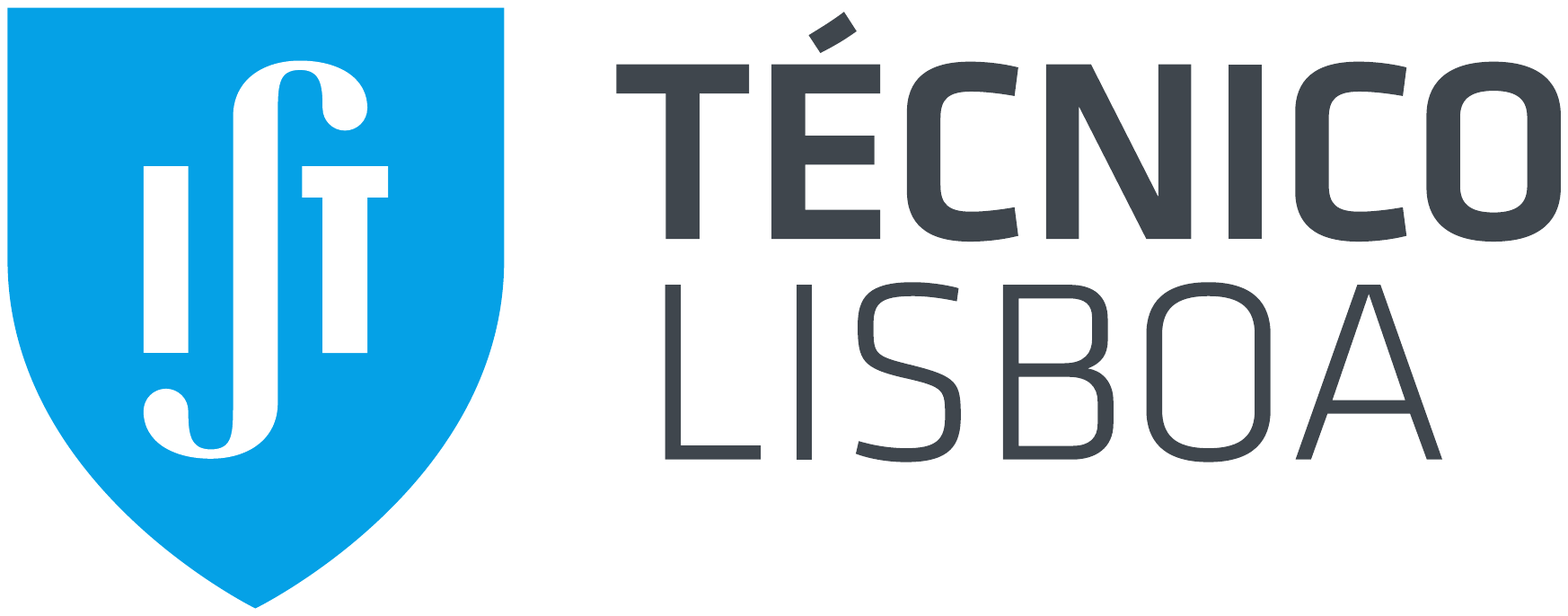}
\\
\\
\begin{center}
%
\vspace{2.5cm}

\vspace{1.0cm}
{\FontLb Superradiant amplification by stars and black holes} \\
\vspace{2.7cm}
{\FontMb João Luís de Figueiredo Rosa} \\
\vspace{2.0cm}
{\FontSn Thesis to obtain the Master of Science Degree in} \\
\vspace{0.3cm}
{\FontLb Physics Engineering} \\
\vspace{1.1cm}
{\FontSn Supervisor: Professor Doutor Vítor Manuel dos Santos Cardoso} \\
\vspace{1.1cm}
{\FontMb Examination Committee} \\
\vspace{0.3cm}
{\FontSn %
\begin{tabular}{ll}
Chairperson: & Professor Doutor José Pizarro de Sande e Lemos \\
Supervisor: & Professor Doutor Vítor Manuel dos Santos Cardoso \\
Member of the Committee: & Professor Doutor Ruben Maurício da Silva Conceição
\end{tabular} } \\
\vspace{1.5cm}
{\FontMb July 2015} \\
\end{center}

\cleardoublepage



\null\vskip5cm%
\begin{flushright}
     To the memory of Líria Guerra
\end{flushright}
\vfill\newpage

\cleardoublepage



\section*{\acknowledgments}

\addcontentsline{toc}{section}{\acknowledgments}

In this section, I want to thank a few people, not only for making this thesis possible but also because they were there for me everytime I needed. These are: the people I worked with and with whom I learned most of what I know about General Relativity; the people I do not work with but with whom I spend everyday, relax, and have fun; and my family who has always been at my side.

To cross the path of GR is one of the most beautiful adventures for a theoretical physicist to live. However, every rose has its thorn, and a huge number of barriers kept appearing in front of me as I dared to walk further into this world. Therefore, the first men I want to thank are the three who not only opened the gates for me but also gave me a hand to walk my first steps into the knowledge of GR: Prof. José Sande Lemos, Dr. Jorge Rocha, and Prof. José Natário.

The help of those men was extremely important for me, but I decided to choose someone else to be my guide as I started to work on this thesis. My next acknowledgement belongs to Prof. Vitor Cardoso, my supervisor, for accepting me as one of his students and giving me the opportunity to start a career in the investigation of GR. Also, I want to give a special thanks to Richard Brito, who very patiently helped me to solve every problem that I faced in the development of this work and who gave me hints and suggestions to proceed. This work would not be possible if it wasn't for his help.

I also want to thank my friends because I believe one can not live a happy life solely based on the success of his carrer. I am satisfied about my work, but I wouldn't be happy if it wasn't for every minute I spent with my friends, specially the members of "Proletariado Internacional" group, Miguel Pinto, and a few others that I do not dare to name because I fear I might forget someone. Thank you all for every night and day out, every trip to the beach or to the mountains, every hike or bike riding, every football match or kart race, and group dinners. 

A huge and special thanks goes to Beatriz Godinho, for every moment we spend togheter, for her endless fountain of support and energy to make me happy, for being always ready to give everything she has to help me solve my problems even if the solution was out of her knowledge, for being patient and a good listener everytime I needed someone to talk to. Thank you for being such an important part of my life, and for teaching me what it is like to share my life and happiness with someone else.

A final but not less important thanks belongs to my family, my parents and my sister, who have always been supportive and whose sacrifices were the main reason for me to be where I am today. Thank you for teaching me that with hard work and the pleasure to live for the reason we love most are the main ingredients for a life of success and happiness. A special thanks to my grandmother who passed away earlier this year and whose dream was to watch me finishing my degree and becoming a successful scientist.

\cleardoublepage



\section*{Resumo}

\addcontentsline{toc}{section}{Resumo}

Nesta tese é estudado o fenómeno da superradiância e as suas implicações na estabilidade de buracos negros e estrelas de fluido perfeito. 

A superradiância é um processo de amplificação de radiação que ocorre em sistemas dissipativos com rotação. Em buracos negros, foi provado que a superradiância existe devido à dissipação no horizonte de eventos, e está associada a fenómenos interessantes, nomeadamente órbitas flutuantes e instabilidades em buracos negros. A superradiância em buracos negros é um tópico muito interdisciplinar, e o seu estudo permite obter resultados importantes na àrea da física de partículas. O espalhamento de um campo escalar por um buraco negro em rotação leva à formação de estados quase-ligados. Quando em rotação, estes estados podem levar a instabilidades relacionadas com a superradiância. Estes resultados foram usados recentemente para impôr limites à massa de partículas fundamentais e candidatos a matéria escura.


Neste trabalho, mostra-se que, quando a dissipação é incluída apropriadamente, é possível obter resultados semelhantes em outros sistemas auto-gravitantes que não buracos negros. É também demonstrado que os efeitos relativistas relacionados com o fenómeno do arrastamento de referenciais são negligíveis neste cálculo.


Parte dos resultados obtidos no decorrer desta tese encontram-se na Ref.~\cite{jl}.

\vfill

\textbf{\Large Palavras-chave:} Buraco-Negro, Superradiância, Campo Escalar, Fluido Perfeito, Dissipação.

\cleardoublepage


\section*{Abstract}

\addcontentsline{toc}{section}{Abstract}

In this thesis we study the phenomenon of superradiance and its implications to the stability of black-holes (BH) and perfect-fluid stars.

Superradiance is a radiation enhancement process that involves rotating dissipative systems. In BH spacetimes, superradiance is due to dissipation at the event horizon, with interesting associated phenomena, namely floating orbits and BH-bombs. BH superradiance is a very interdisciplinary topic, and its study allows us to obtain important results in the area of particle physics. The scattering of a scalar field by a rotating BH leads to the formation of quasi-boundstates. In rotational systems, these states can give rise to superradiant instabilities. These results were recently used to impose constraints to the mass of fundamental particles and darkmatter candidates.


In this work, it is shown that, when dissipation is properly included, similar results are achievable in self-gravitating systems other than black-holes, such as perfect fluid stars. It is also demonstrated that the relativistic effects related to the phenomenon of frame-dragging are neglectable in this calculation.


Some of the results obtained in the development of this thesis can be found in Ref.~\cite{jl}.

\vfill

\textbf{\Large Keywords:} Black-Hole, Superradiance, Scalar Field, Perfect Fluid, Dissipation.

\cleardoublepage



%
\tableofcontents
\cleardoublepage 

%
\listoftables
\addcontentsline{toc}{section}{\listtablename}
\cleardoublepage 

%
\listoffigures
\addcontentsline{toc}{section}{\listfigurename}
\cleardoublepage 

%
%
%
%
%
%
%

\nomenclature[ra]{$a$}{Angular momentum per unit mass}
\nomenclature[rG]{$G_{ab}$}{Einstein tensor}
\nomenclature[rg]{$g$}{Metric determinant}
\nomenclature[rg]{$g_{ab}$}{Metric tensor}
\nomenclature[rJ]{$J$}{Angular momentum}
\nomenclature[rL]{$\mathcal{L}$}{Lagrangian density}
\nomenclature[rP]{$\mathcal{P}_l^m$}{Associate Legendre polynomial}
\nomenclature[rl]{$l,m$}{Eigenvalues of the Spherical harmonics}
\nomenclature[rM]{$M$}{Total mass of the black-hole or star}
\nomenclature[rm]{$m\left(r\right)$}{Mass inside a spherical surface of radius $r$ centered at the origin}
\nomenclature[rP]{$P$}{Pressure of the fluid}
\nomenclature[rR]{$R$}{Ricci scalar, star radius}
\nomenclature[rR]{$R_{ab}$}{Ricci tensor}
\nomenclature[rR]{$R\left(r\right)$}{Radial wave function of the scalar field}
\nomenclature[rr]{$r$}{Radial variable in spherical coordinates}
\nomenclature[rr]{$r_+$}{Radius of the event-horizon}
\nomenclature[rr]{$r_*$}{Tortoise coordinate}
\nomenclature[rS]{$S$}{Action}
\nomenclature[rS]{$S_{ml}$}{Spheroidal harmonic}
\nomenclature[rS]{$S\left(\theta\right)$}{Angular wave function of the scalar field}
\nomenclature[rT]{$T_{ab}$}{Stress-energy tensor}
\nomenclature[rt]{$t$}{time}
\nomenclature[rU]{$U^a$}{Quadrivelocity vector of the fluid}
\nomenclature[rV]{$V\left(r\right)$}{Potential barrier that confines the scalar field}
\nomenclature[ry]{$y_n, j_n$}{Spherical Bessel functions}
\nomenclature[rZ]{$Z$}{Reflection coefficient}
\nomenclature[rZ]{$Z_l^m$}{Spherical harmonic}

\nomenclature[g]{$\alpha$}{Dissipation coefficient}
\nomenclature[g]{$\delta$}{Variation of a variable in the variational principle}
\nomenclature[g]{$\delta_{ij}$}{Kronecker delta, $\delta_{ii}=1, \delta_{ij}=0, i\neq j$}
\nomenclature[g]{$\rho$}{Density of the fluid}
\nomenclature[g]{$\theta$,$\phi$}{Angular variables in spherical coordinates}
\nomenclature[g]{$\omega$}{Frequency of the scalar field}
\nomenclature[g]{$\Omega$}{Angular speed of the event horizon or star surface}
\nomenclature[g]{$\Gamma^a_{bc}$}{Christoffel symbol}
\nomenclature[g]{$\Psi$}{Wave function of the scalar field}
\nomenclature[g]{$\lambda$}{Constant of separation in the method of separation of variables}
\nomenclature[g]{$\kappa$}{Coupling constant of the Einstein's equation}
\nomenclature[g]{$\mu$}{Mass of the scalar field}
\nomenclature[g]{$\zeta$}{Angular speed of an inertial frame as seen from an observer at infinity}

\nomenclature[s]{$\infty$}{Limit to infinity}
\nomenclature[s]{$h$}{Limit to the event horizon}
\nomenclature[s]{$0$}{Limit to the origin}


%
\printnomenclature
\addcontentsline{toc}{section}{\nomname}
\cleardoublepage

%
%
%
%
%
%


\glossary{name={\textbf{GR}},description={General Relativity}}

\glossary{name={\textbf{BH}},description={Black-hole}}

\glossary{name={\textbf{KG}},description={Klein-Gordon}}

\glossary{name={\textbf{QBS}},description={Quasi-boundstate}}

\glossary{name={\textbf{QNM}},description={Quasi-normal mode}}

\printglossary
\addcontentsline{toc}{section}{\glossaryname}
\cleardoublepage

%
\setcounter{page}{1}
\pagenumbering{arabic}



\chapter{Introduction}
\label{chapter:introduction}

In this chapter, the motivation to study the phenomenon of superradiance and the main implications in various areas of physics are discussed. The most important milestones in the history of the Theory of General Relativity and Black-Hole (BH) Superradiance are provided, as well as a brief discussion about the recent results on star superradiance in General Relativity.

\section{Motivation}
\label{section:motivation}

The theory of stellar evolution states that stars whose masses are of the order of the sun's mass will collapse and reach a final equilibrium state as a white dwarf or a neutron star. However, for much larger masses, these final states can't be reached and the collapse will continue to such a point that the gravitational forces overcome the internal pressure and stresses~\cite{kippenhahn}. In this case, there is nothing able to stop the contraction. The fate of such a star can be described by the Theory of General Relativity, which predicts that a collapsing star with a mass in the conditions specified above will necessarily contract until all its matter arrives at a singularity~\cite{oppen1939}. The object formed after this contraction is called a BH. In the last decades it has become evident that BHs have a very particular role in physics with strong implications in many different areas of physics, such as astrophysics, particle physics and high energy physics.

The first stellar-mass BH candidate was discovered by the UHURU satellite, an orbital X-ray observatory launched by NASA in 1970. This satellite discovered over 300 X-ray sources~\cite{giac2002}. Further observations of the fluctuations of one of those sources, Cygnus X-1, showed that the energy generation should take place over a small region of about $10^5$ km~\cite{oda1999}. Measurements of the Doppler shift of a nearby star were performed and used to predict the mass of Cygnus X-1, which was soon accepted to be a BH candidate~\cite{bolton1972}. Since then, several other BH candidates were discovered and nowadays it is even believed that most of the galaxies have a supermassive BH at their center~\cite{korm1995}, including our own~\cite{reid2008}. They are also believed to source some of the most powerful events in the Universe, such as gamma-ray bursts or powering Active Galactic Nuclei~\cite{lynden1969} and it's probable that one of the most exciting breakthroughs of this century will be the detection by the next generation of gravitational wave detectors~\cite{LIGO,VIRGO,KAGRA,ET,ELISA} of gravitational radiation coming directly from binaries of BH or neutron stars.

Not less interesting are phenomena occurring due to the interaction of spinning BHs with generic bosonic fields, related to BH \emph{superradiance}. Superradiance is a radiation enhancement process that occurs in dissipative systems. This process occurs in BHs due to dissipation at the event horizon, which allows for energy and angular momentum to be extracted from the vacuum~\cite{penrose1971} and consequent amplification of the scattered radiation. If confined, these superradiant modes will lead to an amplification \emph{ad infinitum}, giving rise to instabilities like the BH-bomb~\cite{teukolsky1972b}. Such confinement occurs, for example in the presence of massive bosonic fields, where the mass term plays the role of an effective potential barrier for superradiant modes. In particle physics the study of these instabilities has important applications to the search of dark matter candidates and physics beyond the Standard Model~\cite{arvanitaki2010}. These studies were, for example, used to constrain the mass of ultralight degrees of freedom, such as the photon~\cite{vitor1} and the graviton~\cite{vitor2}. Superradiance is also associated, in Einstein's theory coupled to massive complex scalar fields, to the existence of asymptotically flat "hairy" BH solutions~\cite{herd2014}.

BH-bombs are also intresting in other frames of work like the gauge/gravity duality\cite{maldacena1997}. This duality allows us to map the geometry of a BH near the horizon to a two dimensional conformal field theory from which the Bekenstein Entropy can be deduced~\cite{bredberg2011}. Furthermore, extra-dimensional models predict that gravity becomes strong at small distances, but bigger than the 4-dimensional Plank scale, which would allow BHs to be created in particle accelerators~\cite{dimopoulos2001}. BHs may also have the key for the development of a consistent quantum theory of gravity. In fact, Hawking discovery that BHs radiate~\cite{hawking1975}, a semi-classic phenomenon which predicts the evaporation of a BH by emission of thermal radiation, was the first phenomenon to be predicted making use of both general relativity and quantum mechanics.

In this work we shall annalyse the stability and superradiance of a Kerr BH against a massive scalar perturbation. We will show that, when taking into account dissipation in stars, these systems also display superradiant phenomena with implications for their stability. Throughout this thesis we use units where $G=c=\hbar=1$

\section{State-of-the-art}
\label{section:state}

The term ``black-hole'' was used for the first time in 1967 by Wheeler, who defined it as the resultant object of a complete gravitational collapse, which produces a curvature in spacetime so strong that no light or matter can be ejected~\cite{wheeler1971}. Wheeler also stated that any object that falls into a BH loses all its identity except for three parameters: mass, electric charge and angular momentum. This conjecture is known as the No-hair property~\cite{gravitation} and is the reason why BHs are the simplest macroscopic objects in the universe. These objects can be described by the Theory of General Relativity.

Developed by Albert Einstein in 1915, the Theory of General Relativity provides a description of gravity as a geometrical property of spacetime. This description states that spacetime is not flat, like it was assumed in special relativity and Newton's universal law of gravitation, but rather curved by the presence of matter and energy~\cite{einstein1916}. This interaction is described by a set of partial differential equations for the metric tensor called the Einstein's field equations, namely
\be
G_{ab}\equiv R_{ab}-\frac{1}{2}Rg_{ab}=8\pi T_{ab},
\ee
where the Ricci tensor $R_{ab}$ and the Ricci scalar $R=g^{ab}R_{ab}$ are purely geometrical entities which can be obtained from the metric tensor $g_{ab}$, and $T_{ab}$ is the stress-energy tensor associated with the presence of matter. Finding a solution to these equations consists of finding the metric tensor $g_{ab}$ for a given $T_{ab}$. When Einstein first presented his theory, he didn't actually believe that these equations could be solved to find an exact solution. However, in less than one year, the first exact solution was found, by Karl Schwarzschild, which described the gravitational field generated by a static mass point~\cite{schwarz1916}:

\be
ds^2=-\left(1-\frac{2M}{r}\right)dt^2+\left(1-\frac{2M}{r}\right)^{-1}dr^2+r^2\left(d\theta^2+\sin^2\theta d\phi^2\right).
\ee

One can interpret this solution as a vacuum solution exterior to some spherically symmetrical body of radius $R>2M$. There is also a different metric solution to describe the body for $r<R$, which is related to a non-zero energy-stress tensor $T_{ab}$. For $R<2M$, we are in the conditions stated in the first paragraph of the introduction, and a BH is formed. 

The Schwarzschild solution has two singularities, one at $r=2M$ and another at $r=0$. The singularities can be coordinate singularities, which correspond to a weakness in the coordinate system in use and can be removed by performing a coordinate change to another system; or real singularities, which do not depend on the coordinate system in use and represent a region of infinite space curvature. It can be shown that the singularity at $r=2M$ is a coordinate singularity which is a null hypersurface dividing the manifold into two disconnected components. However no coordinate system can remove the singularity at $r=0$, and therefore this is a real or essential singularity~\cite{dinverno}.  

The implications of the Schwarzschild solution were only understood several years later. In 1957 Tulio Regge and John Archibald Wheeler used a perturbative method to analyze the stability of the Schwarzschild geometry and found that this solution is linearly stable~\cite{wheeler1957}. This work marks the beginning of BH perturbation theory, an extremely useful method that is going to be used further in this work. In the following year, the so-called ``golden age of General Relativity'' begins. David Finkelstein performed a coordinate change to the Schwarzschild metric and showed that the hypersurface $r=2M$ is not a singularity but a horizon, that is a unidirectional membrane: causal influences can cross it but only in one direction~\cite{finkel1958}. This result was a huge progress in understanding the solutions of GR.

It took another five years for Roy Patrick Kerr to discover, in 1963, a new solution of Einstein's in vacuum equations describing not a static point particle but a rotating mass~\cite{kerr1963}. The Kerr BH is described, in Boyer-Lindquist coordinates, by the following line element:
\be\label{kerr1}
ds^2=-\frac{\Delta}{\rho^2}\left(dt-a\sin^2\theta d\phi\right)^2+\frac{\sin^2\theta}{\rho^2}\left[\left(r^2+a^2\right)d\phi-adt\right]^2+\frac{\rho^2}{\Delta}dr^2+\rho^2d\theta^2,
\ee
where $a=J/M$ is the angular momentum per unit mass and the parameters $\Delta$ and $\rho$ are given by
\be\label{kerr2}
\Delta=r^2+a^2-2Mr,\ \ \ \ \ \rho^2=r^2+a^2\cos^2\theta.
\ee
Notice that this line element reduces to the Schwarzschild line element if we set the angular momentum to zero, which means that this solution is consistent with the previous one. However it cannot describe the exterior solution of a rotating star, because one cannot match it to an interior solution. This solution has an extreme importance in astrophysics since BHs are formed by collapsing stars, and every star has rotational motion. 

The Kerr solution has two horizons that correspond to the roots of $\Delta$. These horizons correspond to the values $r=M\pm\sqrt{M^2-a^2}$. One can also show that the coordinates $r$ and $\theta$ are not the usual spherical coordinates, by setting the mass of the BH to zero and noticing that the line element obtained is the Minkowski line element in ellipsoidal coordinates. Therefore, our real singularity, that corresponds to $r=0$ and $\theta=\pi/2$ for any $\phi$, has the shape of a ring of radius $a$ in the equatorial plane~\cite{dinverno}. This solution also presents a region outside the horizon, called the ergoregion, where a particle is necessarily propelled in motion with the rotating spacetime. This region is located at the roots of $g_{tt}$, that is $r_{erg}=M+\sqrt{M^2-a^2\cos^2\theta}$.

The understanding of singularities was improved in 1969 by Stephen Hawking and Roger Penrose, who developed the Hawking-Penrose singularity theorems. Singularities can be spacelike, like in the Schwarzschild metric, meaning that the geodesics that cross the event horizon fall into the singularity in a finite time interval; or timelike, meaning that there are null geodesics which can not be extended infinitely to the past, like in the Kerr metric~\cite{hawking1973}. These theorems state that singularities are unnavoidable for some energy conditions and global properties of spacetime. In the collapse of ordinary matter to form a BH, a singularity will always form when a horizon forms, which prevents us from seeing it. This result is known as the Cosmic Censorship Conjecture~\cite{penrose1969}. 

In 1969, Penrose theorised a process to extract energy from a rotating BH. This process is made possible because the rotational energy of the Kerr BH is located outside the horizon in the ergoregion. The so-called Penrose process consists of splitting a particle that falls into the ergoregion in two. We can arrange the momentum of both particles so that one of them falls to the outer horizon of the BH with a negative energy as seen by an observer at infinity, while the escaping particle has a greater energy than the originally infalling one~\cite{penrose1971}. The realization that it was possible to extract energy from a BH, was one of the key steps for the discovery of the main topic of this thesis: \emph{BH superradiance}.  

Robert Henry Dicke was the first to use the term ``superradiance'', when describing the amplification of radiation due to the coherence of the emitters in a quantum system~\cite{dicke1954}. Although not directly related, a similar radiation enhancement process was shown to occur when scattering waves with rotating BHs. This was discovered in 1971 by Yakov Borisovich Zeldovich, who showed that the scattering of incident waves in a rotating dissipative body results in waves with a larger amplitude, if the frequency $\omega$ of the incident wave satisfies the condition
\be\label{super_condition}
\omega<m\Omega,
\ee
where $m$ is the azimuthal number with respect to the rotation axis and  $\Omega$ is the angular speed of the body~\cite{zeldovich1971a}. In the following year, Zeldovich realized that at the quantum level, this would lead rotating bodies to spontaneously emit quantum pairs of particles in the superradiant regime~\cite{zeldovich1971b}. Inspired by this work, in 1975 Hawking used quantum field theory to prove that BHs are thermodynamic bodies, which create and emit particles in the form of thermal radiation. This causes a slow decrease in its mass and eventually causes its total evaporation~\cite{hawking1975}. This was the first result that linked general relativity, quantum mechanics and thermodynamics.

The rotating BH described by the Kerr metric also presents this kind of superradiance. Saul Teukolsky made use of the tetrad formalism developed by Penrose and Ezra Newman in 1962 which treats general relativity in terms of spin coefficients~\cite{newman1962} to decouple and separate the equations for generic bosonic perturbations of the Kerr geometry and obtain a general master equation to describe scalar, electromagnetic and gravitational perturbations. To obtain a separable equation, he assumed that the perturbation wave function could be written in the form~\cite{teukolsky1972}
\be\label{ansatz}
\Psi=e^{-i\omega t}e^{-im\phi}S\left(\theta\right)R\left(r\right)\,,
\ee
and obtained a pair of master equations for the radial function $R\left(r\right)$ and angular function $S\left(\theta\right)$.  Later that year, Teukolsky and William Press solved these equations for a scattering problem and found out that, when the superradiance condition~\eqref{super_condition} is met, the wave inside the potential barrier reinforces the reflected wave on the outside, and the energy of the outgoing wave is greater than the initial one~\cite{teukolsky1974}. This extra energy was shown to come from the rotational energy of the BH, which spins down in the process. 

One equation that remained to be separated was the Dirac equation, which describes fermionic perturbations. This separation was performed in 1973 by William George Unruh who, working independently from Teukolsky, separated this equation for a massless fermion and proved that, unlike the case of bosonic fields, these perturbations do not give rise to superradiance~\cite{unruh1973}. Three years later, this result was generalized to massive fermions, by Subrahmanyan Chandrasekhar~\cite{chandra1976}. Therefore, superradiance is absent for fermionic perturbations, but it is present for scalar, electromagnetic and gravitational perturbations.

BH superradiance is associated with very interesting phenomena~\cite{teukolsky1972b}. Still in 1972, Press and Teukolsky considered the possibility of using a rotating BH to build a BH bomb. If we could confine a rotating BH by building a mirror around it, superradiance would create an instability in the system. For bosonic perturbations, the number of modes exponentially increases in each reflection, and the pressure of radiation in the mirror would increase until it explodes. Another phenomenon also proposed by Press and Teukolsky, was the existence of \emph{floating orbits}. A particle in a stable circular orbit around the BH with an angular speed in the superradiant condition~\eqref{super_condition} emits both radiation to infinity and, at the same time, extracts energy from the BH through superradiance. If at a given radius the particle extracts more energy than it radiates, it will spiral outwards to a new radius where the energy rates are balanced. In this floating radius, the particle will gradually extract energy from the BH, which causes the lowest stable orbit radius to increase until it equals the floating radius. When this happens, the floating ceases and the particle falls into the BH. 

The separation of the perturbation field equations was a huge step which allowed to study the linear stability of the Kerr BH. Given that this metric is believed to describe \emph{all} BHs in the Universe, this was by far one the most important achievements of the BH ``golden age''. To study the stability of a BH against a perturbation, one must compute the frequency $\omega$ (see eq. \eqref{ansatz}) which is an eigenvalue of the radial equation with certain boundary conditions. These conditions are: at infinity, the wave must be outgoing, because we discard unphysical waves coming from infinity; and at the horizon, the wave must be ingoing, because by definition nothing can cross the horizon in the outwards direction, and it can also be shown that the outgoing solution is not regular and must be discarded. The result is a complex frequency describing how the BH responds to a small perturbation. If the imaginary part of this frequency is negative, then the scalar field decreases exponentially with time (see \eqref{ansatz}), and the BH is stable against that perturbation. The BH will then vibrate with specific frequencies which are called the Quasi-normal modes (QNMs). The term ``quasi'' refers to the fact the system is dissipative, and thus small vibrations decay in time. 

The Kerr BH was shown to be linearly stable against massless bosonic perturbations~\cite{teukolsky1972b}, but Steven Detweiler, showed that the Kerr BH is unstable against massive scalar perturbations with an azimuthal number $m>0$, due to the BH bomb effect~\cite{detweiler1980}. In fact, the scalar mass term acts as a mirror and is able to confine superradiant modes leading to the instability. More recently, Kerr BHs were shown to be linearly unstable against massive vector field perturbations~\cite{vitor1} and massive tensor field perturbations~\cite{vitor2}. Since these fields can describe massive photons and gravitons respectively, the observation of spinning BHs was used to derive upper bounds for the mass of these particles. The bounds obtained were $m_\nu < 10 \times 10^{-22}$ eV and $m_g < 5 \times 10^{-23}$ eV. These are extremely tiny masses, far beyond what we can detect in particle accelerators.

The existence of the superradiant instability also led to the construction of asymptotically flat hairy BH solutions~\cite{hod2012}\cite{herd2014}. These solutions are thought to be one of the possible end-states of superradiant instabilities for massive complex scalar fields: the Kerr BH suffers a phase transition to a hairy BH at the superradiance threshold. They can be thought of as solutions describing Kerr BHs surrounded by scalar clouds. These were first obtained at the linear level by Shahar Hod in 2012~\cite{hod2012}, while full non-linear solutions were obtained numerically by Carlos Herdeiro and Eugen Radu in 2014~\cite{herd2014}. They are of extreme importance since they go against the no-hair theorem, which has been playing a very important role in the development of BH physics.

Last but not least, BH superradiance was also studied at the microscopic level by making use of the AdS/CFT duality, developed by Juan Maldacena in 1997, a conjecture mapping two apparently different theories: gravitational theories in the $d$-dimensional Anti-de Sitter spacetime and conformal field theories (CFT) in $d-1$ dimensions~\cite{maldacena1997}. The related theories live in spaces with a different number of dimensions. In fact, the dual CFT space has less one dimension than AdS, and so this conjecture is said to be holographic: CFT lives at the boundary of AdS. In 1999 the AdS/CFT correspondence was used by Irene Bredberg, Thomas Hartman, Wei Song and Andrew Strominger to describe BH superradiance for a nearly-extreme BH in terms of a dual two-dimensional conformal field theory in which the BH corresponds to a thermal state and the scalar field corresponds to a specific operator~\cite{bredberg2010}. Also relevant for the AdS/CFT conjecture, five years later, Vítor Cardoso and Óscar Dias showed that Kerr-AdS BHs are unstable because the AdS boundary provides a natural confinement mechanism for superradiant radiation~\cite{vitor3}.

BH superradiance is still an open and timely topic of research. For example, superradiance was shown to work at a full nonlinear level only very recently~\cite{east2014}. Even at the linear level, superradiant instabilities of generic massive bosonic perturbations are still research topics. The discovery of hairy solutions associated to these instabilities, opened the possibility for the development of new and exciting physics. 

It is now clear that the phenomenology of BH superradiance is extremely rich, which might lead us to ask if these results also apply to other self-gravitating systems such as stars or neutron stars. Since ideal stars do not dissipate energy, they also do not display superradiance and none of the associated instabilities, except for possible ergoregion instabilities~\cite{reviewvitor}. However, real stars do dissipate energy and, as shown by Zeldovich~\cite{zeldovich1971b} in 1971, any classical system with dissipation should give rise to superradiant scattering. Stars in full General Relativity were recently shown to display superradiance as long as dissipation is included~\cite{richartz2013}. The main objective of this thesis is to demonstrate quantitatively that a perfect-fluid constant density star displays superradiance, as well as instabilities against massive perturbations, and also that the relativistic effects of frame-dragging are neglectable in the calculation of the amplification factors.

\cleardoublepage



\chapter{Black-hole superradiance}
\label{chapter:introduction}

We will be working with the Kerr solution of the Einstein equations, since the Schwarzschild solution can be obtained from this one. The Kerr metric is given, in Boyer-Lindquist coordinates, by \eqref{kerr1} and \eqref{kerr2}. The radius of the event horizon can be obtained by finding the roots of $\Delta$. There are two solutions for this equation, and the event horizon corresponds to the greater one, $r_+=M+\sqrt{M^2-a^2}$. Notice that if $a>M$, this radius is no longer a real number, and therefore we define the extreme Kerr BH with $a=M$. In this work we shall only deal with BH solutions with angular momenta below this limit.

The equation that describes a scalar field $\Psi$ is the Klein-Gordon (KG) equation. For more information on how to deduce this equation see appendix \ref{anexo}. We are also going to use Teukolsky's ansatz~\cite{teukolsky1972} for the wave function in order to obtain an equation that is separable in the radial and angular coordinates. This ansatz is simply a Fourier transform of the form \eqref{ansatz}.

\section{The wave equation}

Inserting the previous ansatz into the KG equation, expanding and separating the sum inside the derivative, and substituting the inverse of the metric, the exponential terms cancel in both sides of the equation and it becomes separable in $r$ and $\theta$. Dividing through by $R\left(r\right)S\left(\theta\right)$, using the fact that $\sin^2\theta=1-\cos^2\theta$ and separating the variables we obtain the following equations:
\be
\omega^2a^2\cos^2\theta-2ma\omega-\frac{m^2}{\sin^2\theta}+ \frac{1}{S}\partial_\theta\partial_\theta S+\frac{1}{S}\cot\theta\partial_\theta S  - \mu^2a^2\cos^2\theta=\lambda,
\ee
for the angular equation, and:
\be
\omega^2a^2-\frac{1}{\Delta}\left(a^2+r^2\right)^2-\frac{1}{\Delta}4Mrma\omega-\frac{1}{\Delta}m^2a^2-\frac{1}{R}{{\Delta}'}\partial_rR-\frac{\Delta}{R}\partial_r\partial_rR+\mu^2r^2=\lambda,
\ee
for the radial equation, where a prime denotes a differentiation in order to the only variable of a certain function, and $\lambda$ is the constant of separation. Now that the equations are separated, we can multiply the angular one by $S\left(\theta\right)$ and the radial one by $R\left(r\right)/\Delta$ in order to obtain two linear second order differential equations of the form
\beq\label{rad}
&&{R}''\left(r\right)+A_r\left(r\right) {R}'\left(r\right)+B_r\left(r\right) R\left(r\right)=0,\nonumber \\
&&{S}''\left(\theta\right)+A_\theta\left(\theta\right) {S}'\left(\theta\right)+B_\theta\left(\theta\right) S\left(\theta\right)=0,
\eeq
where the coefficients are given by
\be
A_r\left(r\right)=\frac{{\Delta}'}{\Delta},\ \ \ A_\theta\left(\theta\right)=\frac{\cos\theta}{\sin\theta},
\ee
\be
B_r\left(r\right)=-\frac{\mu^2r^2}{\Delta}+\omega^2\frac{\left(a^2+r^2\right)^2}{\Delta^2}+\frac{4Mrm\omega a}{\Delta^2}+m^2\frac{a^2}{\Delta^2}+\frac{\omega^2a^2}{\Delta}+\frac{\lambda}{\Delta};\nonumber
\ee
\be
B_\theta\left(\theta\right)=-\left(\mu^2+\omega^2\right)a^2\cos^2\theta-\frac{m^2}{\sin^2\theta}-\lambda.
\ee

These are the angular and radial equations that we are going to solve for the static and rotating black-hole solutions. 

\subsection{The angular equation}

Imposing $a=0$, i.e. for the Schwarzschild solution, the angular equation simplifies to
\be
{S}''+\frac{\cos\theta}{\sin\theta}{S}'+\left(-\frac{m^2}{\sin^2\theta}+\lambda\right)S=0.
\ee

In order to solve this equation, notice that multiplying through by $\sin\theta$ makes it possible to write the first two terms as a product derivative. Then, multiplying again by $\sin\theta$ and dividing by $S\left(\theta\right)$ we can write the equation in the useful form
\be\label{sphe}
\frac{\sin\theta}{S}\partial_\theta\left(\sin\theta\partial_\theta S\right)+\lambda\sin^2\theta=m^2,
\ee
which corresponds to the equation for the spherical harmonics. The Sturm-Liouville problem imposes the separation constant to take the form $\lambda=l\left(l+1\right)$, and the solution is the associate Legendre polynomial $\mathcal{P}_l^m$
\be
S\left(\theta\right)=N\mathcal{P}_l^m\left(\cos\theta\right),
\ee
where $N$ is an arbitrary constant. Writing the angular part of our wave equation as $Z_l^m\left(\theta,\phi\right)=e^{-im\phi}S\left(\theta\right)$ and using the orthonormality relations for the Legendre polynomials and for the spherical harmonics,
\be
\int_{-1}^{1}\mathcal{P}_l^m\left(x\right)\mathcal{P}_{{l}'}^m\left(x\right)dx=\frac{2\left(l+m\right)!}{\left(2l+1\right)\left(l-m\right)!}\delta_{l{l}'},
\ee
\be
\int_\Omega Z_l^m\left(\theta,\phi\right) Z_{{l}'}^{*{m}'}\left(\theta,\phi\right) d\Omega =\delta_{m{m}'}\delta_{l{l}'},
\ee
we can compute the constant
\be
N=\left[\frac{\left(2l+1\right)\left(l-m\right)!}{4\pi\left(l+m\right)!}\right]^{\frac{1}{2}}.
\ee

If we again consider the general case, we obtain a very similar equation given by
\be
\frac{\sin\theta}{S}\partial_\theta\left(\sin\theta\partial_\theta S\right)+\left(\lambda+a^2\left(\omega^2-\mu^2\right)\cos^2\theta\right)\sin^2\theta=m^2,
\ee
which corresponds to the equation for the spheroidal harmonics. Notice that in the Schwarzschild limit $a=0$ we recover equation \eqref{sphe}. The only difference between this case and the static one is that the Legendre polynomials become the spheroidal harmonics $S_{ml}\left(-ic,\cos\theta\right)$, where $c^2=a^2\left(\omega^2-\mu^2\right)$. These functions are tabulated (see e.g.~\cite{abramovich}). In this case, the constant of separation is $\lambda=l\left(l+1\right)+\mathcal{O}\left(c\right)$~\cite{paolo2013}. In this work we will study quasi-boundstates, which are formed when $\omega\sim\mu$. In this case, $c\sim 0$ and the spheroidal harmonics can be approximated by the spherical harmonics. This problem does not exist in slowly rotating stars because in this case the wave equation is completely separable with the usual spherical harmonics, and therefore $\lambda=l\left(l+1\right)$.

\subsection{The radial equation}

The radial equation is harder to interpret in the form that it was obtained. In order to obtain a more useful form of this equation, we shall perform a change in the radial function and in the radial coordinate. The new radial function and coordinate (known as the tortoise coordinate) are given by
\be
u\left(r\right)=\sqrt{a^2+r^2}R\left(r\right),\ \ \ \ \ f\left(r\right)=\frac{dr}{dr^*}=\frac{\Delta}{r^2+a^2}.
\ee
Thus, a straightforward calculation consisting of computing the derivatives in order to the new variable leads to a new radial equation for the new variables $u$ and $r^*$ in the form of a wave equation in the presence of a potential barrier,
\be
\frac{d^2u}{dr^{*2}}+\left[\omega^2-V\left(r\right)\right]u=0,
\ee
where the potential is given by
\be
V\left(r\right)=f\left(r\right)\left[\frac{1}{r^2+a^2}\left(\mu^2r^2-\omega^2a^2+\frac{4Mram\omega}{\Delta}-\frac{m^2a^2}{\Delta}+\lambda\right)+\frac{\Delta+{\Delta}'r}{\left(r^2+a^2\right)^2}-\frac{3r^2\Delta}{\left(r^2+a^2\right)^3}\right],
\ee

Since the potential is a complicated function of the radial coordinate $r$, we can't write an analytical solution that holds for the entire space. The radial function must be computed numerically by imposing appropriate boundary conditions at the horizon and at the infinity. These boundary conditions may be written as a combination of ingoing and outgoing waves in the form of complex exponentials:
\be
\label{equ}
u\left(r_i\right)=A_i e^{+i\omega_ir^*}+B_i e^{-i\omega_ir^*},
\ee
where $A_i$ and $B_i$ are constants and the subscript $i=h,\infty$ represents the limits to the horizon and to the infinity. To find the effective frequencies, let's first of all compute the limits of the function $f\left(r\right)$:
\be
\lim_{r\to\infty}f\left(r\right)=1,\ \ \ \ \ \ \lim_{r\to r_{+}}f\left(r\right)=0.
\ee
Using the first result we can solve the limit of $V\left(r\right)$ for $r\to\infty$ and obtain the effective frequency by
\be
\lim_{r\to\infty}V\left(r\right)=\mu^2\ \implies\ \omega_\infty=\sqrt{\omega^2-\mu^2}.
\ee

The horizon limit is a bit more complicated since the use of the limit of $f\left(r\right)$ multiplied by the terms in $V\left(r\right)$ either leads to a trivial result or an indetermination $0/0$. The only terms that do not result in a limit equal to zero are the ones which are divided by $\Delta$, and therefore those are the only terms that have to be considered at the horizon. The potential then becomes
\be
\lim_{r\to r_{+}}V\left(r\right)=\frac{1}{\left(r_+^2+a^2\right)^2}\left(4Mr_+am\omega-m^2a^2\right).
\ee
To simplify this limit, notice that at the horizon $\Delta=0$ and therefore we can write the denominator of this function as $r_+^2+a^2=2Mr_+$. Also, if we sum and subtract $\left(2Mr\omega\right)^2$ inside the parenthesis, we can find a perfect square and write our potential in the form
\be
V\left(r_{+}\right)=\omega^2-\left(\omega-\frac{ma}{2Mr_+}\right)^2.
\ee
The term $\omega^2$ cancels out with the one in the wave equation, therefore our effective frequency takes the final form
\be
\omega_h=\omega-m\Omega,\ \ \ \ \ \Omega=\frac{a}{2Mr_+},
\ee
where $\Omega$ is the angular speed of the event horizon.

Now, we expect that $A_h=0$, since the horizon works as a one directional membrane and therefore the outgoing wave solution should not be regular. To show that, we perform a change to the time and azimuthal coordinates of the form
\be
v=t+\int\frac{r^2+a^2}{\Delta}dr,\ \ \ \ \phi^*=\phi-\int\frac{a}{\Delta}dr,
\ee
to the Teukolsky ansatz, from which we obtain a new radial function of the form
\be
R^*=e^{-im\int\frac{a}{\Delta}dr+i\omega\int\frac{r^2+a^2}{\Delta}dr}R.
\ee

A straightforward calculation reveals that, since the radial function before the coordinate change is proportional to exponential functions of the form $e^{\pm\left(\omega-m\Omega\right)r_*}$, the new radial function is then
\be
R^*\sim e^{i\int\frac{1}{\Delta}\left(r^2+a^2\right)\left[\left(\omega-m\Omega\right)\pm\left(\omega-m\Omega\right)\right]},
\ee
where we have used the definition of $\Omega$ written above. Since at the horizon $\Delta\to 0$, the numerator inside the integral must also tend to zero in order for this radial function to be regular. Therefore, we are obligued to choose the coefficient $A_h=0$. Also, in order to normalize the amplitude of the wave at the horizon to 1, we impose $B_h=1$. Note that this choice is completely arbitrary due to the linearity of the wave equation. The parameters $A_\infty$ and $B_\infty$ are unknown and shall remain as free parameters until otherwise stated.

\section{Numerical results}

In this section we will develop a numerical method to solve the radial equation. This method is extremely useful and allows us to study the existence of superradiance, the quasi-boundstates (QBS) and the stability of the Kerr BH against scalar perturbations (similar methods can be used to study the exact same processes but with a different perturbation, such as vectorial or tensorial.)

Solving this equation by numerical methods leads to some computational problems. First of all, we can't start the numerical integration exactly at the horizon because, since $\Delta\to 0$, the radial function is not stable at this point. Also, it is impossible to perform a numerical integration all the way up to infinity. To solve these two problems, we define two parameters: a small parameter $\epsilon<<1$ to be summed to the horizon radius, and a "numerical infinity" to stop the integration, which we normally set to be a function of the frequency. In order to conclude that we have found a solution of our equations, we have to verify that small changes on these parameters do not affect the results.

These two parameters solve the problem of the integration limits but impose another one, since the boundary conditions that we obtained are only valid at the horizon and at infinity. To solve this problem, we shall expand these solutions in Taylor series around the horizon and the infinity and use those series as boundary conditions.

\subsection{Boundary conditions}

To find out the series expansions to be used, we shall solve equation \eqref{equ} back to the radial function $R\left(r\right)$ and radial coordinate $r$. First of all, notice that there is an analytical solution to write the tortoise coordinate as a function of $r$,
\be
r^*=r+\frac{M}{\sqrt{M^2-a^2}}\left[r_+\log\left(r-r_+\right)-r_-\log\left(r-r_-\right)\right].
\ee
In the limit $r\to r_+$, the first logarithmic function diverges and therefore we can neglect the other two terms. Thus, taking the factors multiplied outside the logarithm to the exponent of the function inside it, the exponential solution of the wave equation takes the form
\be
R_h\left(r\right)\sim\left(r-r_+\right)^\alpha,\ \ \ \ \alpha=-i\frac{\left(\omega-m\Omega\right)Mr_+}{\sqrt{M^2-a^2}}.
\ee
On the other hand, if we take the limit $r\to\infty$, we can not neglect the term $r$ anymore. In this case, we first separate the two terms into two different exponentials and then apply the previous method to the one containing the logarithmic functions. The result is as follows:
\be
\label{inf1}
R_\infty\left(r\right)\sim e^{i\sqrt{\omega^2-\mu^2} r}r^\beta,\ \ \ \ \beta=i\frac{M\left(2\omega^2-\mu^2\right)}{\sqrt{\omega^2-\mu^2}}.
\ee

Considering the previous limits, we shall perform the following expansions:
\be
R_h\left(r\right)=\left(r-r_+\right)^\alpha\sum_n C_n \left(r-r_+\right)^n,
\ee
\be
\label{inf2}
R_\infty\left(r\right)=e^{\pm i\sqrt{\omega^2-\mu^2} r}r^{\pm\beta}\sum_n D_n \frac{1}{r^n},
\ee
up to a given $n$. These expansions must be inserted into the radial differential equation \eqref{rad} in order to obtain $n-1$ equations for the coefficients $C_n$ and $D_n$ as functions of the coefficients $C_0$ and $D_1$, respectively. Notice that the expansion at infinity starts at $n=1$ because the term $n=0$ leads to an identity when we perform this replacement. The coefficient $C_0$ is simply given by $B_h=1$ by the normalization performed before, and the coefficient $D_1$ is either $A_\infty$ or $B_\infty$ for the outgoing or ingoing wave, respectively.

\subsection{Existence of superradiance}

With this method we aim to demonstrate the existence of superradiance in Kerr BHs. The boundary conditions imposed to the radial function are the series expansion at the horizon and its derivative. When the numerical integration is computed for a given $\omega$, we obtain a radial function that depends on the coefficients $A_\infty$ and $B_\infty$. These coefficients can be computed by matching the solution with the boundary conditions at infinity.

The coefficients computed are complex numbers. The reflection coefficient can be computed simply by the absolute value of the outgoing wave coefficient divided by the absolute value of the ingoing wave coefficient, 
\be\label{amp}
Z=\frac{|A_\infty|^2}{|B_\infty|^2}.
\ee

\begin{figure}[h!]
\begin{minipage}{.5\textwidth}
\centering
\includegraphics[width=8cm]{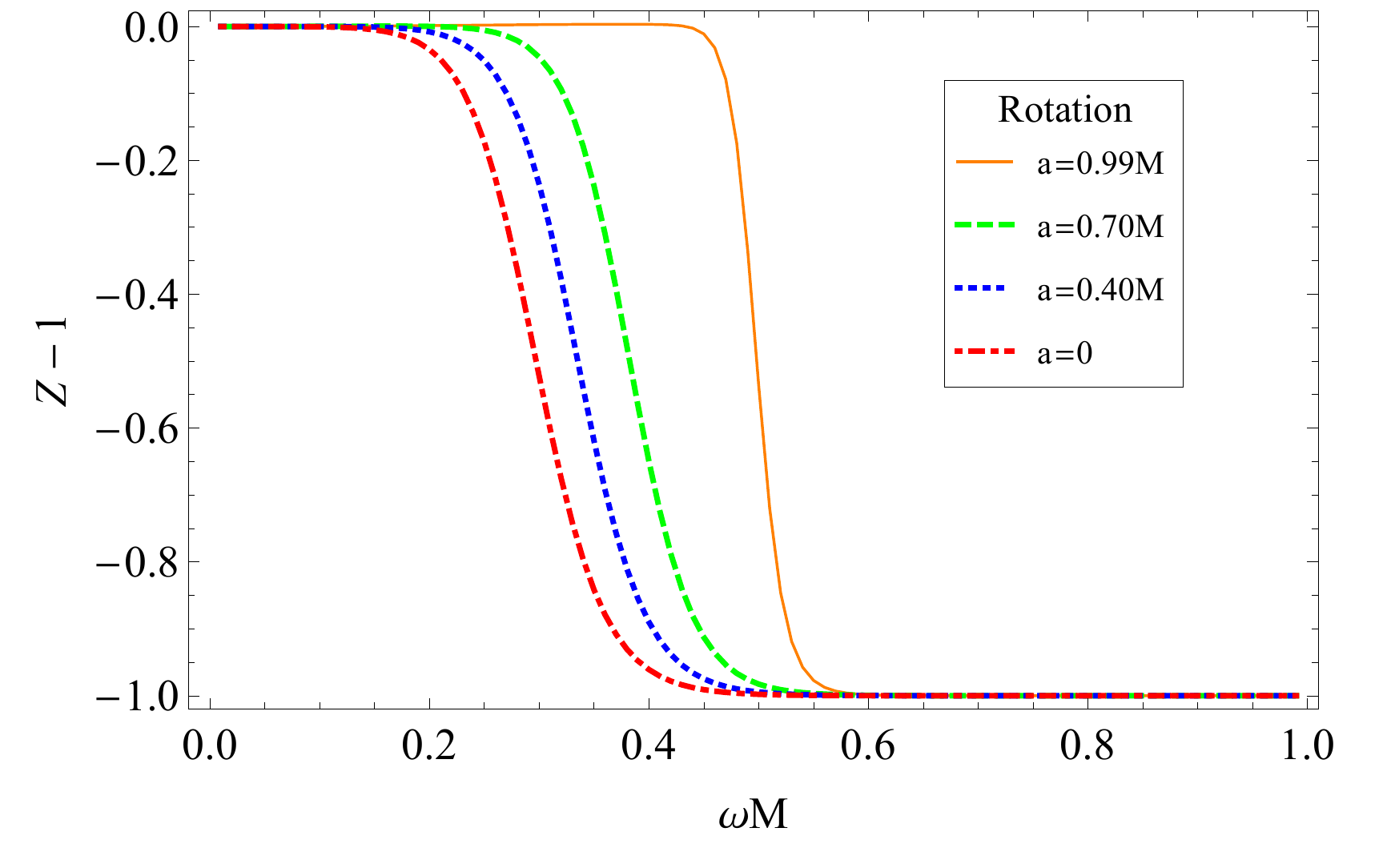}
\caption{Reflection coefficient in linear scale for $\mu=0$, $l=m=1$}
\label{superlinmu}
\end{minipage}%
\ \ \ \ \
\begin{minipage}{.5\textwidth}
\centering
\includegraphics[width=8cm]{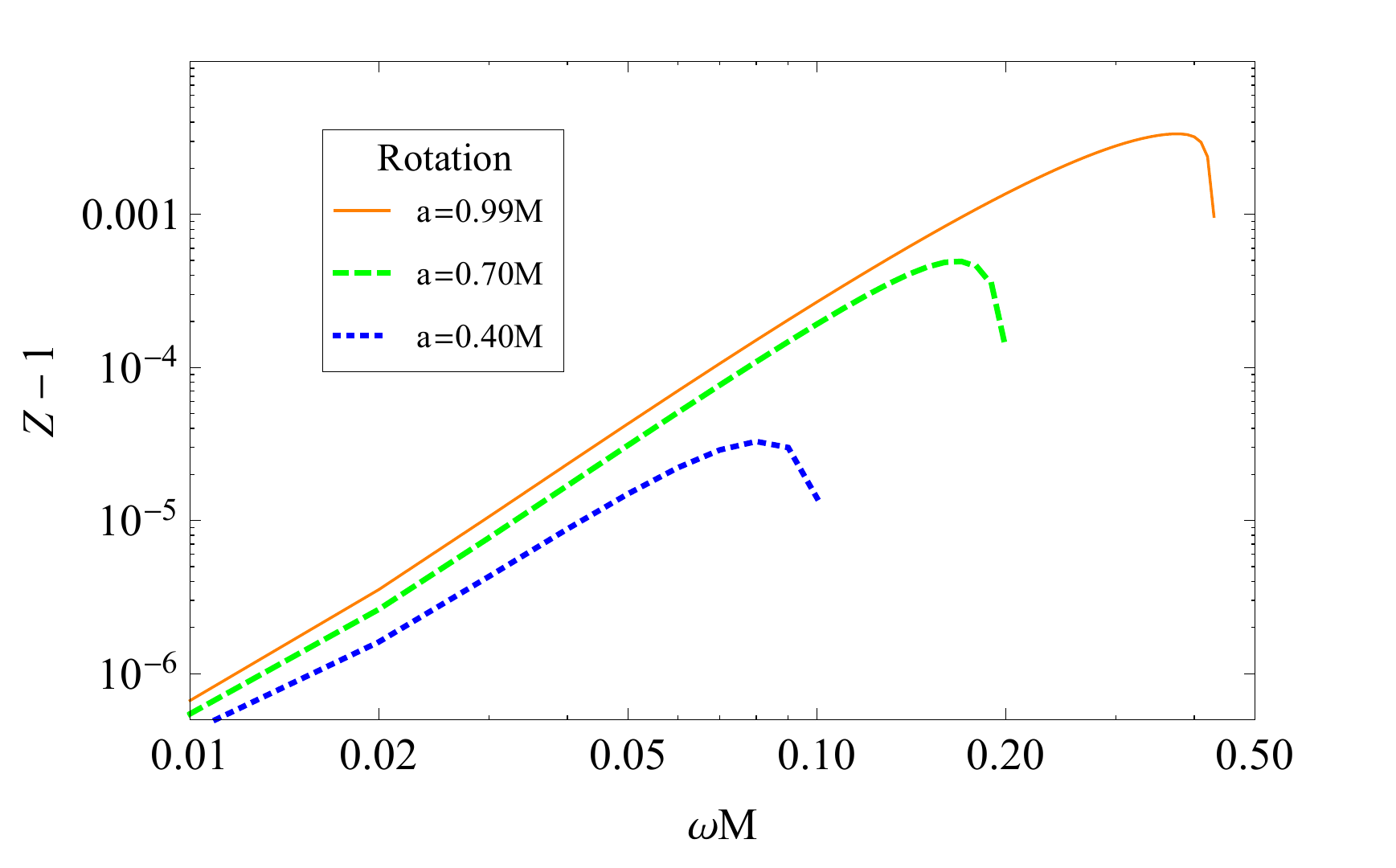}
\caption{Reflection coefficient in logarithmic scale for $\mu=0$, $l=m=1$}
\label{superlogmu}
\end{minipage}
\end{figure}

By repeating this calculation for various values of $\omega$ it is possible to build a plot of this reflection coefficient as a function of $\omega$ for a massless perturbation, as can be seen in figures \ref{superlinmu} and \ref{superlogmu}. For different values of $a$ the shape of the function $Z\left(\omega\right)$ is similar, differing only on the threshold frequency. However, we show that only the solutions with $a>0$ have a frequency region where the reflection coefficient is greater than one, which means that the outgoing wave as seen from an observer at the infinity has a greater energy than the incident wave. We also verify that the maximum frequency for which $Z=1$ is $\omega=\Omega$, which corresponds to the threshold of the superradiance condition \eqref{super_condition}.

\begin{figure}[h!]
\begin{minipage}{.5\textwidth}
\centering
\includegraphics[width=8cm]{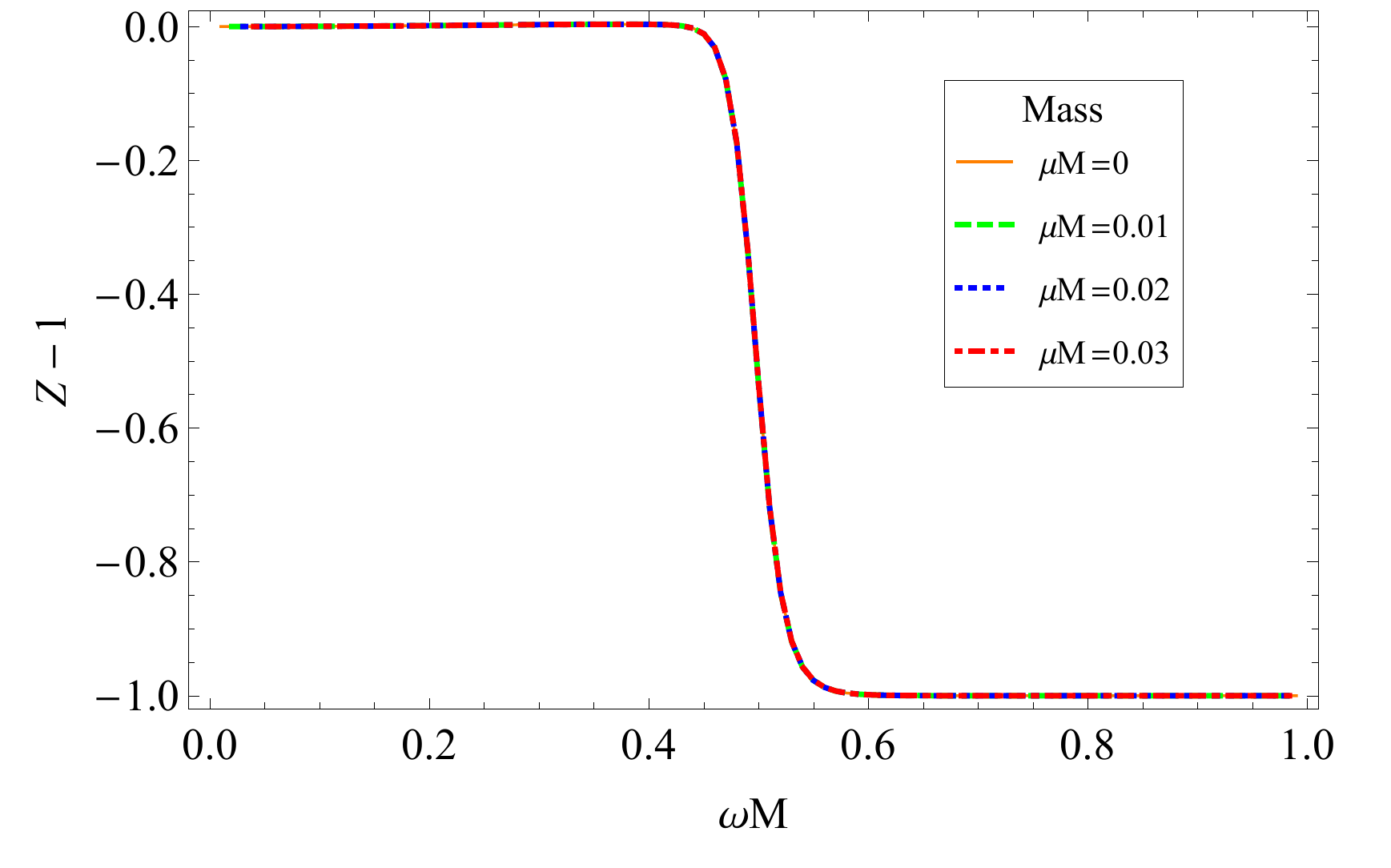}
\caption{Reflection coefficient in linear scale for $a=0.99M$, $l=m=1$}
\label{superlina}
\end{minipage}
\ \ \ \ \
\begin{minipage}{.5\textwidth}
\centering
\includegraphics[width=8cm]{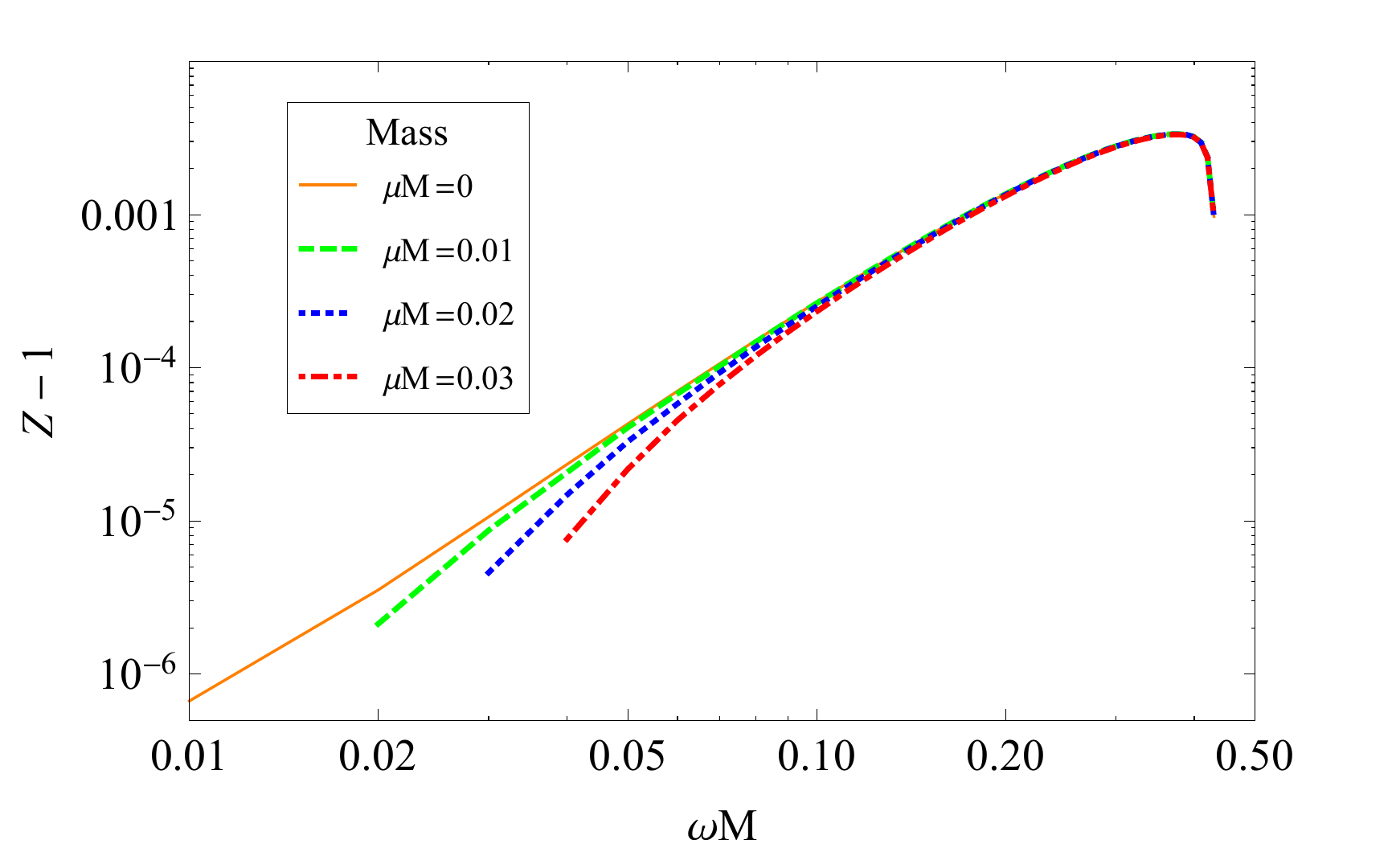}
\caption{Reflection coefficient in logarithmic scale for $a=0.99M$, $l=m=1$}
\label{superloga}
\end{minipage}
\end{figure}

We can also repeat this calculation for a fixed $a$ and varying the mass $\mu$, as done in figures \ref{superlina} and \ref{superloga}. We show shown in this case that the threshold frequency does not change, which is expected since $\Omega$ does not vary. We can also see that for higher values of the mass the amplification coefficient is lower for low frequencies but, as the frequency approaches the threshold frequency, $Z$ approaches the value for a massless perturbation. In this case, we must have $\mu<\omega$ because otherwise we do not have waves.

The same method that we used to compute the QBS could also be used to compute the Quasi-Normal Modes (QNM). However, these vibrational modes are much harder to obtain, and for this method to be successful we had either to perform series expansions to orders above 10~\cite{molina2010} or to search a root using initial values very close to the correct value. There are other semi-analytical methods which can find these vibrational modes very accurately, such as the continued-fraction method developed by Leaver in 1985~\cite{leaver1985}, which is the most accurate method up to date. However, this method can not be applied to dissipative systems other than BHs, which makes it useless in the star superradiance study that we aimed to do.

\subsection{Quasi boundstates}

The quasi boundstates (QBS) are formed when the frequency $\omega$ of a given perturbation is lower than its mass $\mu$. In this situation, our solution at the infinity given by equation \eqref{equ} becomes a combination of real exponentials rather than complex ones. Thus, the wave behaviour in the radial coordinate is lost. It is said that the mass works as a natural confinement to the perturbation around the BH. The boundary condition at the infinity then becomes
\be
u\left(r_i\right)=A_\infty e^{-\sqrt{\mu^2-\omega^2}r^*}+B_\infty e^{+\sqrt{\mu^2-\omega^2}r^*},
\ee
which implies, in order to preserve the regularity of this function at infinity, like we did before at the horizon, that the coefficient $B_\infty$ must be zero.

This method is very similar to the previous one. The radial differential equation is solved subjected to the same boundary conditions at the horizon as before and the coefficients $A_\infty$ and $B_\infty$ that are coherent with the boundary conditions at infinity are computed. However, instead of solving this equation for different values of $\omega$, we defined a function which receives $\omega$ as an input and returns the value of $B_\infty$. In order to obtain a solution with physical significance, we must compute the roots $\omega_0$ for which $B_\infty=0$. These frequencies are the so-called quasi-boundstates (QBS) and are complex numbers of the form $\omega=\omega_R+i\omega_I$. These frequencies are discrete because this is an eigenvalue problem for the frequency $\omega$.

\begin{figure}[h!]
\begin{minipage}{.5\textwidth}
\centering
\includegraphics[width=8cm]{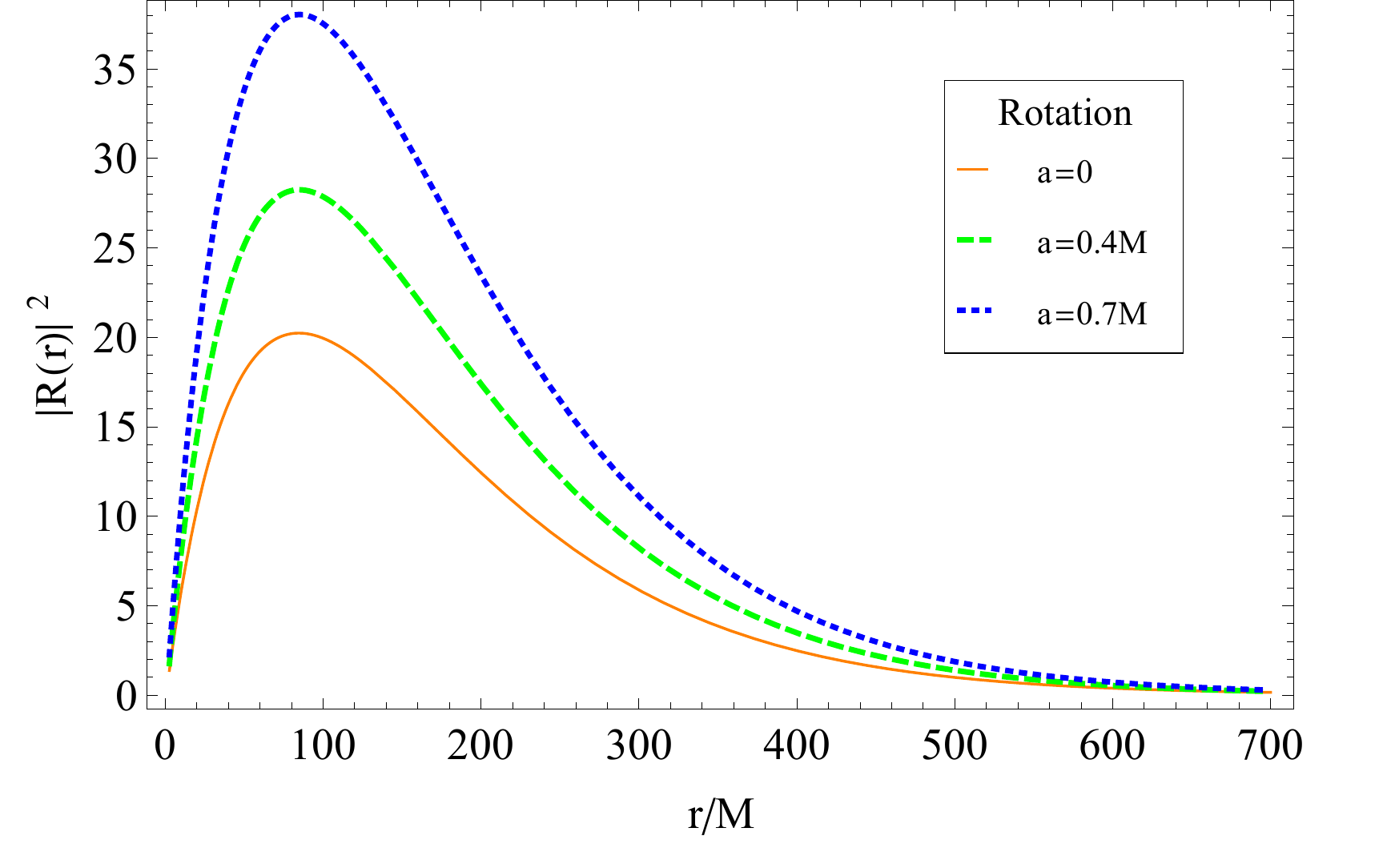}
\caption{Quasi-Boundstates for $\mu M=0.15$, $l=m=1$}
\label{blackqbsa}
\end{minipage}
\ \ \
\begin{minipage}{.5\textwidth}
\centering
\includegraphics[width=8cm]{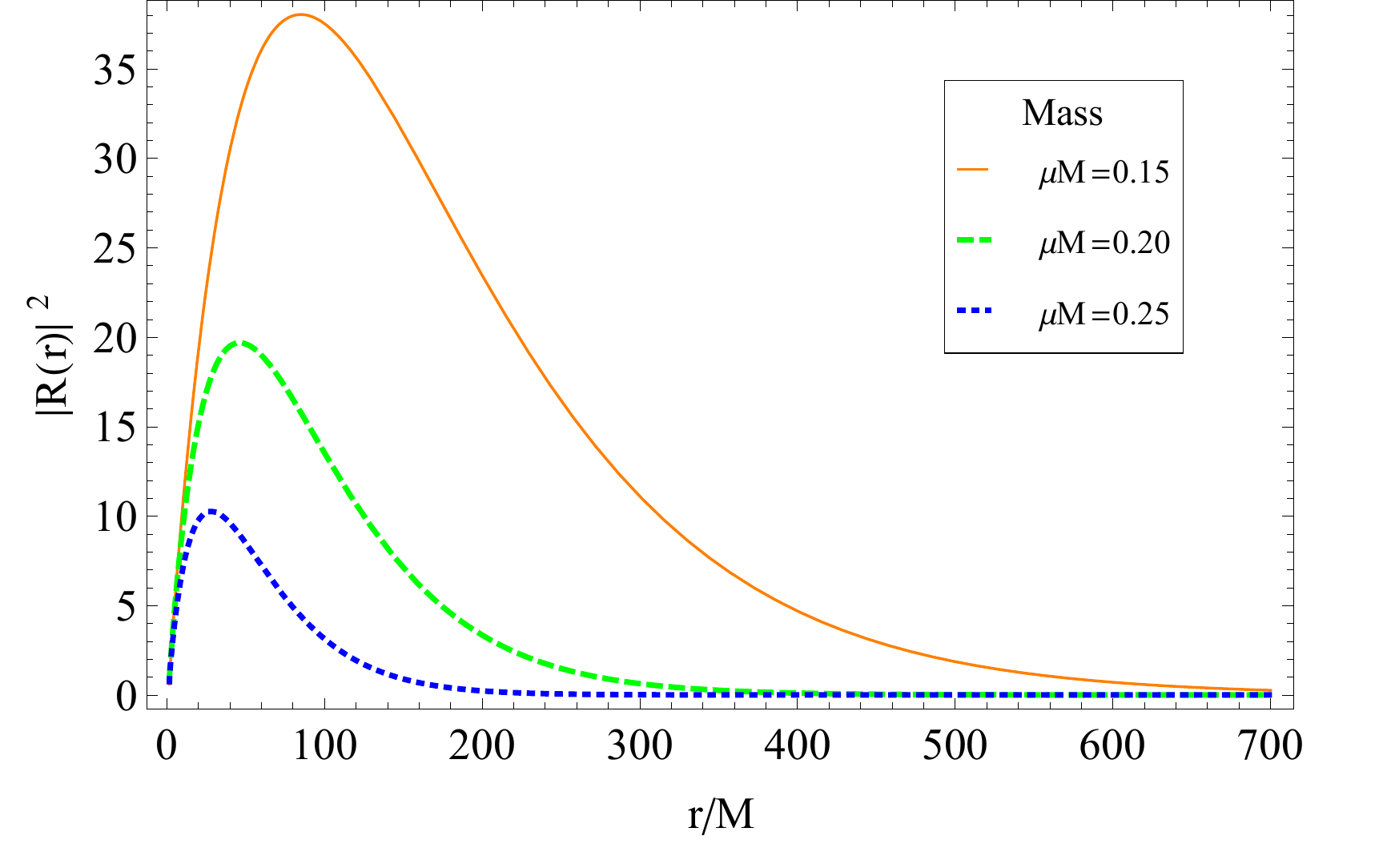}
\caption{Quasi-Boundstates for $a=0.7M$, $l=m=1$}
\label{blackqbsmu}
\end{minipage}
\end{figure}

In figures \ref{blackqbsa} and \ref{blackqbsmu} we show how the quasiboundstates for the fundamental level vary with $a$ and $\mu$. We show that as we increase the angular momentum, the QBS peak remains at the same position. However, if we increase the mass $\mu$, the peak lies closer to the event horizon, which is expected since we know that the peak's position is proportional to $\mu^{-2}$.

We can study the stability of a BH against a perturbation by making use of these QBS. The idea is to rewrite the wave function for the scalar perturbation as
\be
\label{staby}
\Psi=e^{\omega_I t}e^{-im\phi-i\omega_R t} R\left(r\right)S\left(\theta\right).
\ee
We see that if $\omega_I>0$, the scalar field increases exponentially with time. This increase causes the field to stop being a perturbation on the metric, and the system becomes unstable against that perturbation. However, if $\omega_I<0$, the scalar field decays exponentially with time, which means that the perturbation rapidly fades away and the system returns to its original geometry. In this case, the system is stable against that perturbation.

To compute the frequency roots, an initial value of $\omega$ must be set in order to start the numerical process. To find an aproximate value of the real part, we use an analytical result from ref.~\cite{detweiler1980} given by
\be
\label{omegateo}
\omega\sim\mu-\frac{\mu\left(M\mu\right)^2}{2\left(n+l+1\right)},
\ee
which is only valid for $M\mu<<1$, where $n$ is the order of the mode. For the imaginary part of the frequency, since it has a very low value, usually of orders bellow $10^{-6}$, we simply impose the signal of $\omega_I$. Since the instabilities occur for positive values of $\omega_I$ and these instabilites are a consequence of superradiance, then we impose a positive sign if the real part of the frequency, which is aproximately equal to its absolute value, verifies the superradiant condition \eqref{super_condition}.

\begin{figure}[h!]
\begin{minipage}{.5\textwidth}
\centering
\includegraphics[width=8cm]{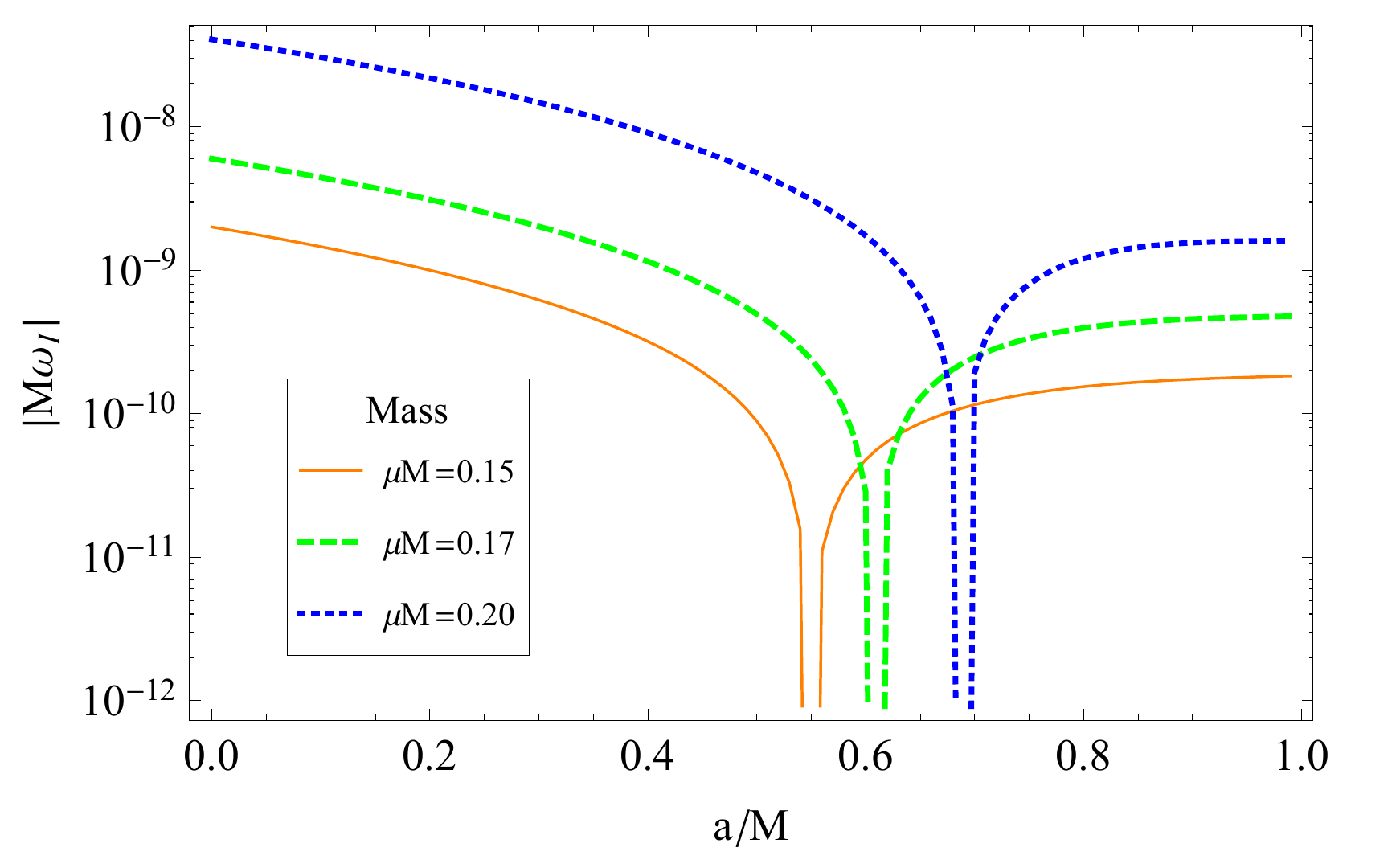}
\caption{Absolute value of the imaginary part of the QBS frequency for $\mu M=0.15, 0.17, 0.20$, $l=m=1$}
\label{blackinstmu}
\end{minipage}
\ \ \ \ \
\begin{minipage}{.5\textwidth}
\centering
\includegraphics[width=8cm]{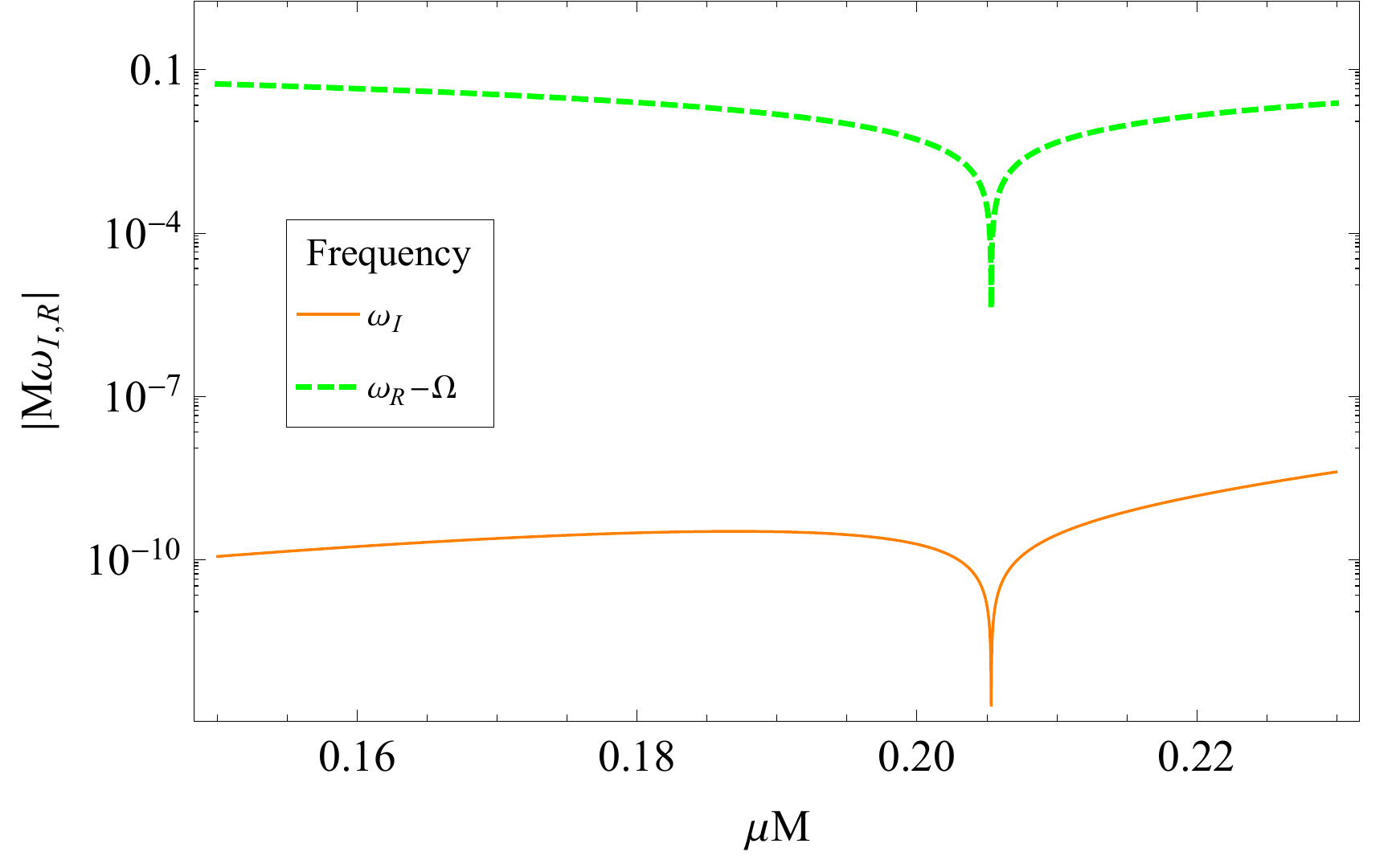}
\caption{Absolute value of the imaginary part of the QBS frequency for $a=0.7M$, $l=m=1$}
\label{blackinsta}
\end{minipage}
\end{figure}

The QBS frequencies were computed for solutions with angular momenta from $a=0$ to $a=1$ for three different field masses. The real part of the frequency is aproximately constant throughout the entire momentum range. These values can be seen in table \ref{wreal}, aside with the analytical values predicted by \eqref{omegateo}. As expected, since this approximation is only valid for low values of $\mu$, the error increases as we increase $\mu$. However, the imaginary part starts from a negative value at $a=0$ and increases to a positive value at $a=1$. For greater masses, the value of $a$ at which $\omega_I$ changes its sign is larger than for lower masses, as can be seen in Figure \ref{blackinstmu}. 

Finally, we obtained a plot of $\omega_I$ and $\omega_R-m\Omega$ as functions of $\mu$, as can be seen in figure \ref{blackinsta}. We verify in this plot that $\omega_I$ starts from a positive value at $\mu=0$ and it inscreases as we increase $\mu$. It is clear that $\omega_I$ changes its sign exactly when $\omega_R=m\Omega$, which means that the stability region occurs when the frequency reaches the threshold frequency, that is, superradiance ceases to exist.
\begin{table}[h!]
\centering
\begin{tabular}{c|c|c|c}
$\mu$ & $\omega_R$ & $\omega_{ana}$ & $\sigma$(\%) \\ \hline
0.15  & 0.149576   & 0.149578     & 0.0013 \\
0.17  & 0.169366   & 0.169386     & 0.0118 \\
0.20  & 0.198953   & 0.199000     & 0.0236 \\ 
\end{tabular}
\caption{Comparison between the real part of the QBS modes obtained numerically and their analytical values. (Valid for M$\mu<<1$)}
\label{wreal}
\end{table}

\cleardoublepage



\chapter{Superradiance in stars}
\label{chapter:introduction}

In this chapter we shall extend the analysis of superradiance to a perfect fluid constant density star. In this case, the metric tensor can be written as~\cite{shapiro}:
\be\label{fluid}
ds^2=-e^{2\varphi}dt^2+\left(1-\frac{2m\left(r\right)}{r}\right)^{-1}dr^2+r^2\left(d\theta^2+\sin^2\theta d\phi^2\right),
\ee
where the mass enclosured in a spherical surface of radius $r$, $m\left(r\right)$, and the exponential function are given by:
\be
m\left(r\right)=\frac{4}{3}\pi r^3\rho,\ \ \ \ \ e^{\varphi}=\frac{3}{2}\sqrt{1-\frac{2M}{R}}-\frac{1}{2}\sqrt{1-\frac{2Mr^2}{R^3}},
\ee
where $M$ is the total mass of the star and $R$ is the radius of the star. The stress-energy tensor $T_{ab}$ that describes a perfect fluid is given by:
\be\label{stress}
T^{ab}=\left(\rho+P\right)U^aU^b+Pg^{ab},
\ee
where $\rho$ is the density of the fluid which is considered to be constant, $P$ is the radial pressure which is a function of the density, and $U^a$ is the quadrivelocity vector. This vector is a timelike unit vector, and therefore it is imposed that $U_aU^a=-1$. It is also known that $U^r=U^\theta=U^\phi=0$ and $U^t=\sqrt{-g^{tt}}$. Finally, it can be shown that the only expression for the pressure that can solve the field equations is:
\be
P=\rho\left(\frac{\sqrt{1-2Mr^2/R^3}-\sqrt{1-2M/R}}{3\sqrt{1-2M/R}-\sqrt{1-2Mr^2/R^3}}\right).
\ee

Dissipation must also be included in the problem, and that can be achieved by the addition of a Lorentz-invariance breaking modification to the Klein-Gordon equation inside the star and in a frame co-rotating with the star as
\be\label{modie}
\nabla_a\nabla^a\Psi+\alpha\frac{\partial\Psi}{\partial t}=\mu^2\Psi.
\ee

Outside the star, the metric reduces to the Schwarzschild metric. The metric \eqref{fluid} describes a static perfect fluid star and therefore, in order to obtain a star that is rotating from the point of view of an observer at infinity, we need to perform a coordinate transformation to a co-rotating frame. This is obtained by performing an azimuthal coordinate change inside the star of the form $\phi'=\phi-\Omega t$. Following the ansatz \eqref{ansatz}, this transformation leads to a new frequency $\omega'=\omega-m\Omega$. It is clear form this expression that in the superradiant regime \eqref{super_condition} the amplification factor $\alpha\omega'$ becomes negative and the medium amplifies radiation, instead of absorbing it.

\section{Newtonian perfect-fluid star}

To simplify our problem, we first consider a star in classical Newtonian gravity. In this section, the relativistic effects related to the phenomenon of frame-dragging are neglected and the star is considered to be rotating in a static spherically symmetric background. The full GR problem shall be approached in the next section.

\subsection{The wave equation}

Once again, inserting the ansatz \eqref{ansatz} into the new Klein-Gordon equation, expanding and separating the sum inside the derivative, and substituting the inverse of the metric, the exponential terms cancel out and we obtain an equation which is separable in $r$ and $\theta$. Dividing through by $R\left(r\right)S\left(\theta\right)r^{-2}$, we obtain the equations:
\be
f\left(r\right)\left[e^{-\varphi}{e^{\varphi}}'+\frac{2}{r}+f^{-1}\left(r\right)\left(\frac{m\left(r\right)}{r^2}-\frac{{m}'\left(r\right)}{r}\right)\right]\frac{r^2}{R}\partial_rR+\omega^2e^{-2\varphi}r^2+f\left(r\right)\frac{r^2}{R}\partial_r\partial_rR-\mu^2r^2-i\omega\alpha r^2=\lambda,\nonumber 
\ee
\be
f\left(r\right)=\left(1-\frac{2m\left(r\right)}{r}\right),
\ee
for the radial equation, and:
\be
-\cot\theta\frac{\partial_\theta S}{S}+\frac{m^2}{\sin^2\theta}-\frac{\partial_\theta\partial_\theta S}{S}=\lambda,
\ee
for the angular equation. Notice that this angular equation is exactly the one obtained in the last chapter in the limit $a=0$, and therefore it shall not be studied again here. The radial equation can be multiplied through by $\frac{R}{r^2}\left(1-\frac{2m\left(r\right)}{r}\right)^{-1}$ in order to obtain again a linear differential equation of the form \eqref{rad}, where the coefficients are given by:
\be
A_r\left(r\right)=e^{-\varphi}{e^{\varphi}}'+\frac{2}{r}+\left(1-\frac{2m\left(r\right)}{r}\right)^{-1}\left(\frac{m\left(r\right)}{r^2}-\frac{{m}'\left(r\right)}{r}\right),
\ee
\be
B_r\left(r\right)=\left(1-\frac{2m\left(r\right)}{r}\right)^{-1}\left(\omega^2e^{-2\varphi}-\mu^2-i\omega\alpha-\frac{\lambda}{r^2}\right).
\ee
Notice that we still need to perform a transformation $\omega\to\omega-m\Omega$.

Just like before, this radial equation has a non-trivial dependence on the radial coordinate and, therefore, the solutions must be obtained numerically. To do so, series expansions are used at infinity and at the origin. Since the metric reduces to the Schwarzschild metric outside of the star, the method of the previous chapter tells us that the series expansion at infinity is the same as before (see \eqref{inf1},\eqref{inf2}). At the origin, keeping only the dominant terms of the radial equation in the limit $r\to 0$, the equation becomes:
\be
{R}''\left(r\right)+\frac{2}{r}{R}'\left(r\right)-\frac{l\left(l+1\right)}{r^2}R\left(r\right)=0,
\ee
where we used $\lambda=l\left(l+1\right)$. The solution for this equation can be obtained by using an ansatz of the form $R\left(r\right)=r^\alpha$, from which we obtain:
\be
R\left(r\right)=A_0 r^l + B_0 r^{-l-1}.
\ee
However, only the solution $r^l$ is regular at the origin because the other one diverges in the limit $r\to 0$, which imposes $B_0=0$. Since the value of $A_0$ is arbitrary, we choose it to be 1. Therefore, we shall perform a series expansion at the origin of the form
\be\label{ori}
R_0\left(r\right)=r^l\sum_n C_n r^{2n},
\ee
where we only consider even powers of $r$ because the coefficients $C_n$ for odd $n$ are identically equal to zero. The coefficient $C_0$ can be set to $A_0=1$ and the parameters $A_\infty$ and $B_\infty$ are once again unknown and shall remain as free parameters until otherwise is stated.

\subsection{Numerical results}

The procedure to compute the amplification factors and quasi-boundtates is very similar to the one used in the previous chapter. However, in this case we have to solve the radial equation separately in the interior and exterior regions. We first integrate the equation from the origin to the surface of the star, and the solution at $r=R$ is used as a boundary condition to integrate from the surface up to the infinity. 

\begin{figure}[h!]
\begin{minipage}{.5\textwidth}
\centering
\includegraphics[width=8cm]{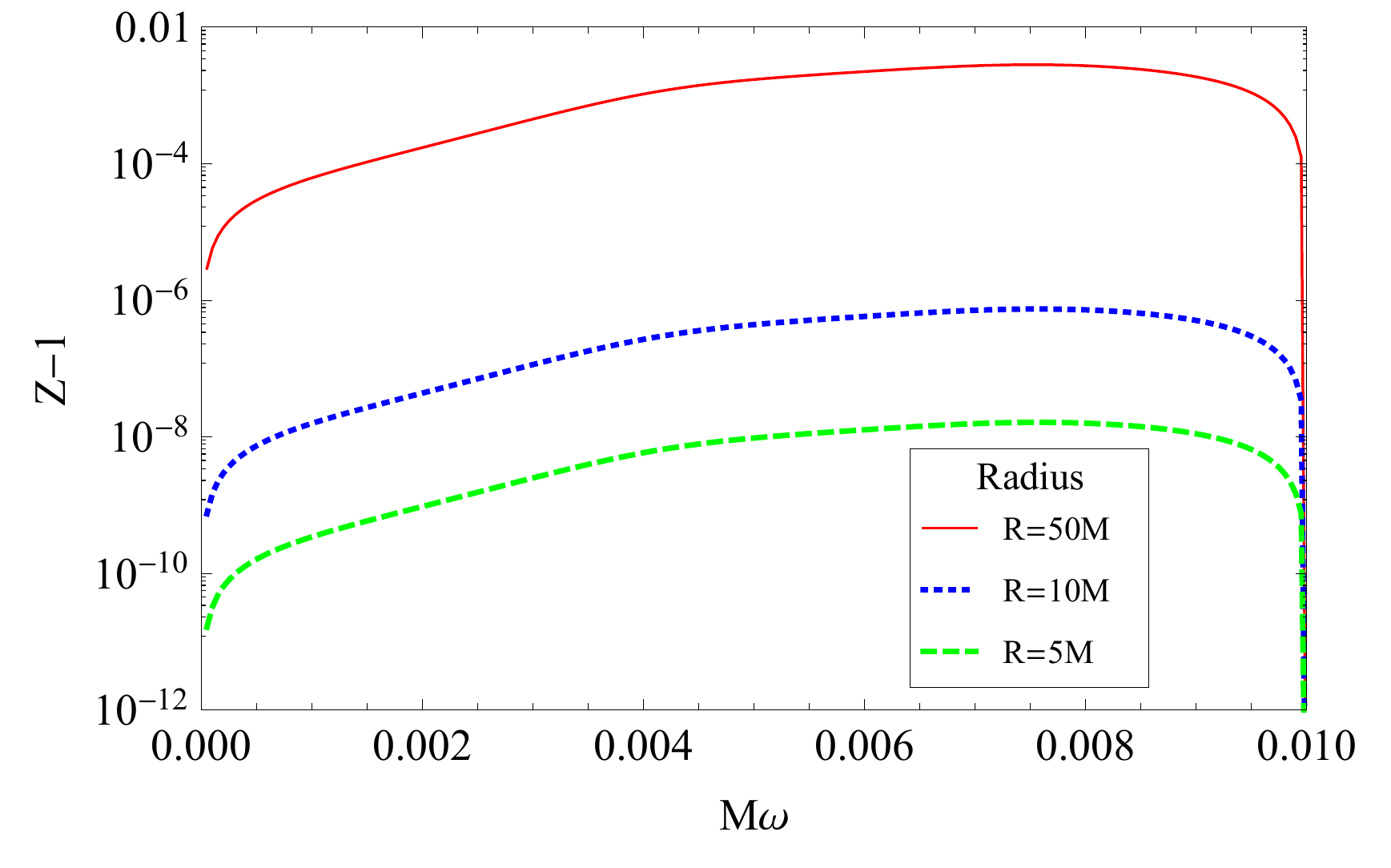}
\caption{Reflection coefficient for $\mu=0$, $\alpha M=0.1$, $\Omega M=0.01$, $l=m=1$.}
\label{superlogR}
\end{minipage}
\ \ \ \ \
\begin{minipage}{.5\textwidth}
\centering
\includegraphics[width=8cm]{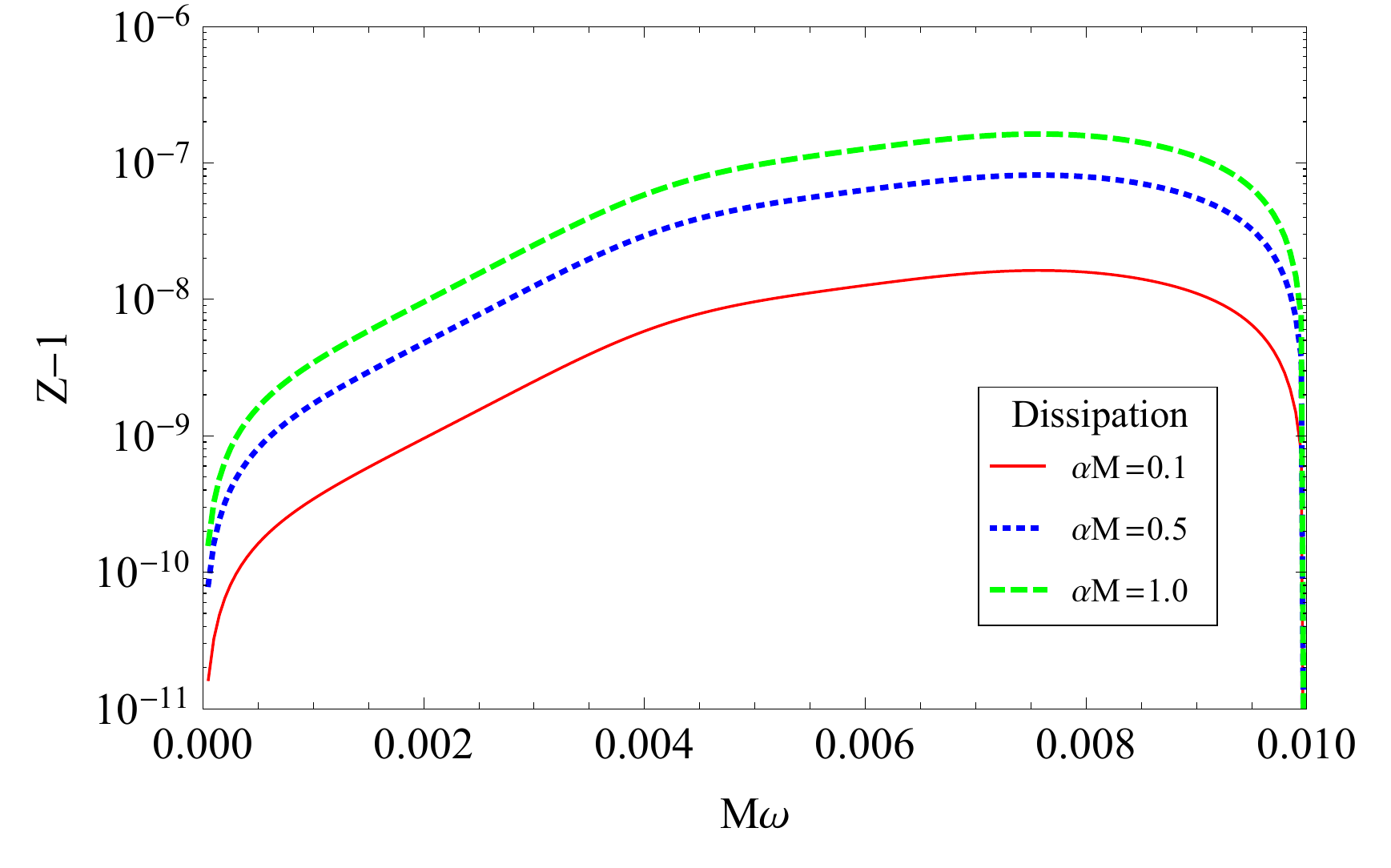}
\caption{Reflection coefficient for $\mu=0$, $R=5M$, $\Omega M=0.01$, $l=m=1$.}
\label{superlogAl}
\end{minipage}
\end{figure}

The amplification factors \eqref{amp} for various values of the star radius were obtained and are shown in figure \ref{superlogR}. For every value of $R$ the shape of $Z\left(\omega\right)$ is similar but the value increases with the radius, which is expected since as we increase the radius of the star for a fixed angular speed we are increasing the linear speed at the surface. We also verify that the threshold frequency occurs in the limits of the superradiant regime \eqref{super_condition}, as expected.

If we instead plot the amplification factors for various values of $\alpha$, as can be seen in figure \ref{superlogAl}, we see that their value increase as we inscrease the dissipation coefficient. This shows that superradiance could not occur without dissipation and therefore it should increase with it. However, we are going to show later that there is a value of $\alpha$ that maximizes the instability factor.

\begin{figure}[h!]
\begin{minipage}{.5\textwidth}
\centering
\includegraphics[width=8cm]{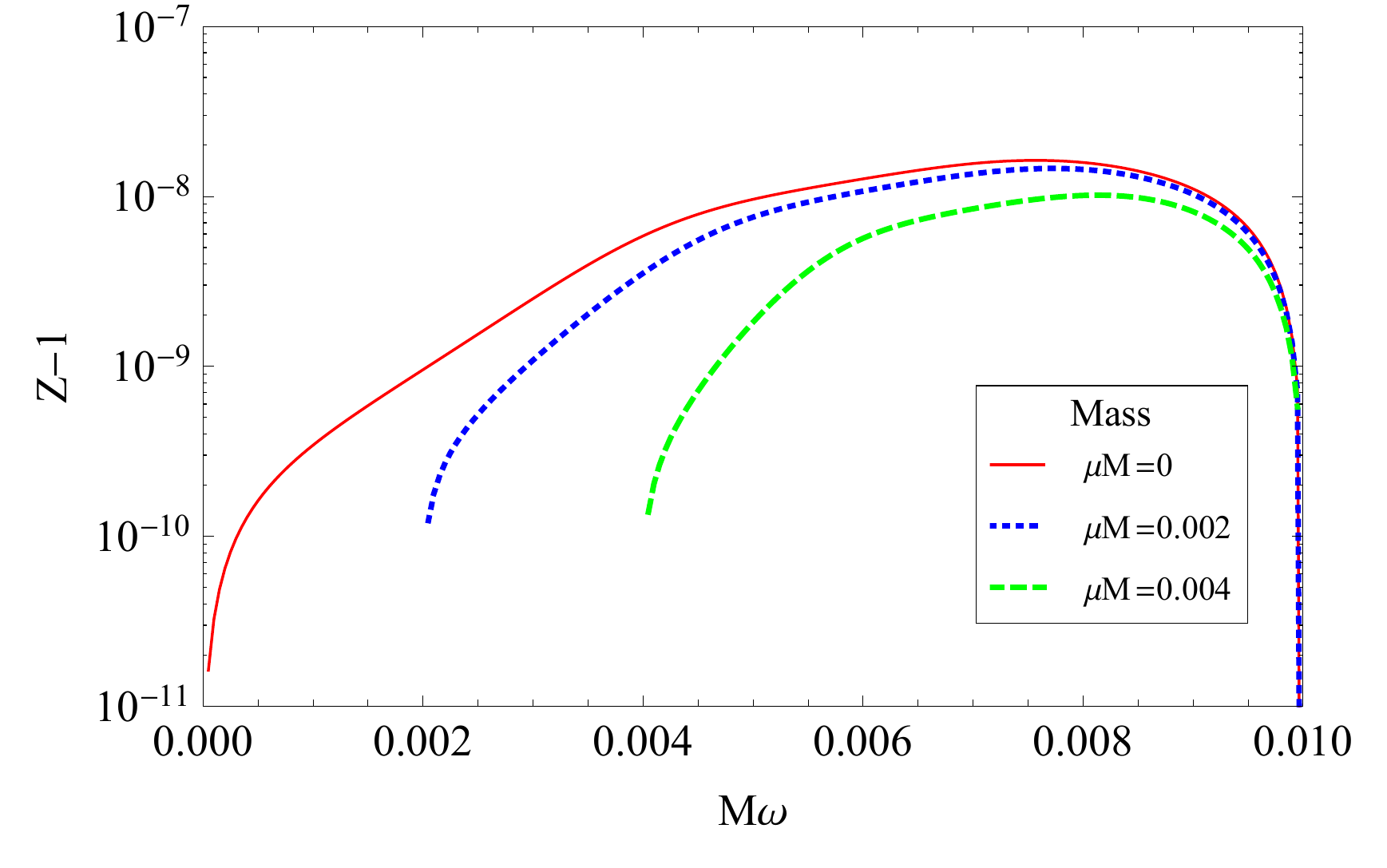}
\caption{Reflection coefficient for $R=5M$, $\alpha M=0.1$, $\Omega M=0.01$, $l=m=1$.}
\label{superlogM}
\end{minipage}
\ \ \ \ \
\begin{minipage}{.5\textwidth}
\centering
\includegraphics[width=8cm]{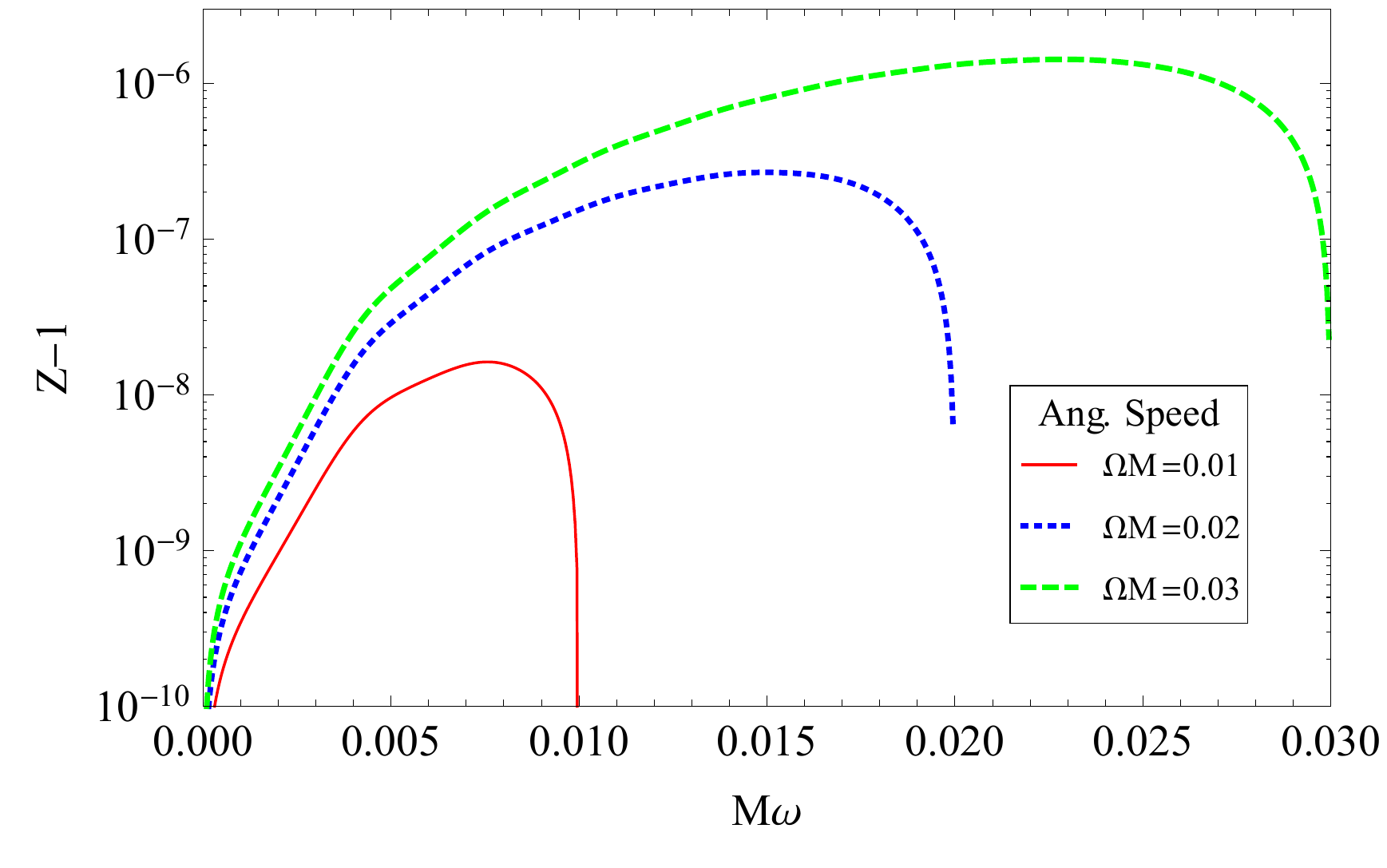}
\caption{Reflection coefficient for $\mu=0$, $\alpha M=0.1$, $R=5 M$, $l=m=1$.}
\label{superlogO}
\end{minipage}
\end{figure}

We can also plot the amplification for different values of $\mu$ like we did in the last chapter. The result is basically the same, for low frequencies the amplification is lower for higher field masses but it approaches the value for a massless perturbation as we get closer to the threshold frequency. Also notice that amplification only occurs in the region $\mu<\omega$, because otherwise there are no wave solutions. These results are shown in figure \ref{superlogM}.

Finaly, in figure \ref{superlogO} we plot the amplification for different values of $\Omega$. The result is also similar to the one obtained in the previous chapter. The threshold frequency increases as we increase $\Omega$, since we alter the limits of the superradiant region. Also, the amplification increases as we increase $\Omega$, which again tells us that this phenomenon is closely related to rotation.

\begin{figure}[h!]
\begin{minipage}{.5\textwidth}
\centering
\includegraphics[width=8cm]{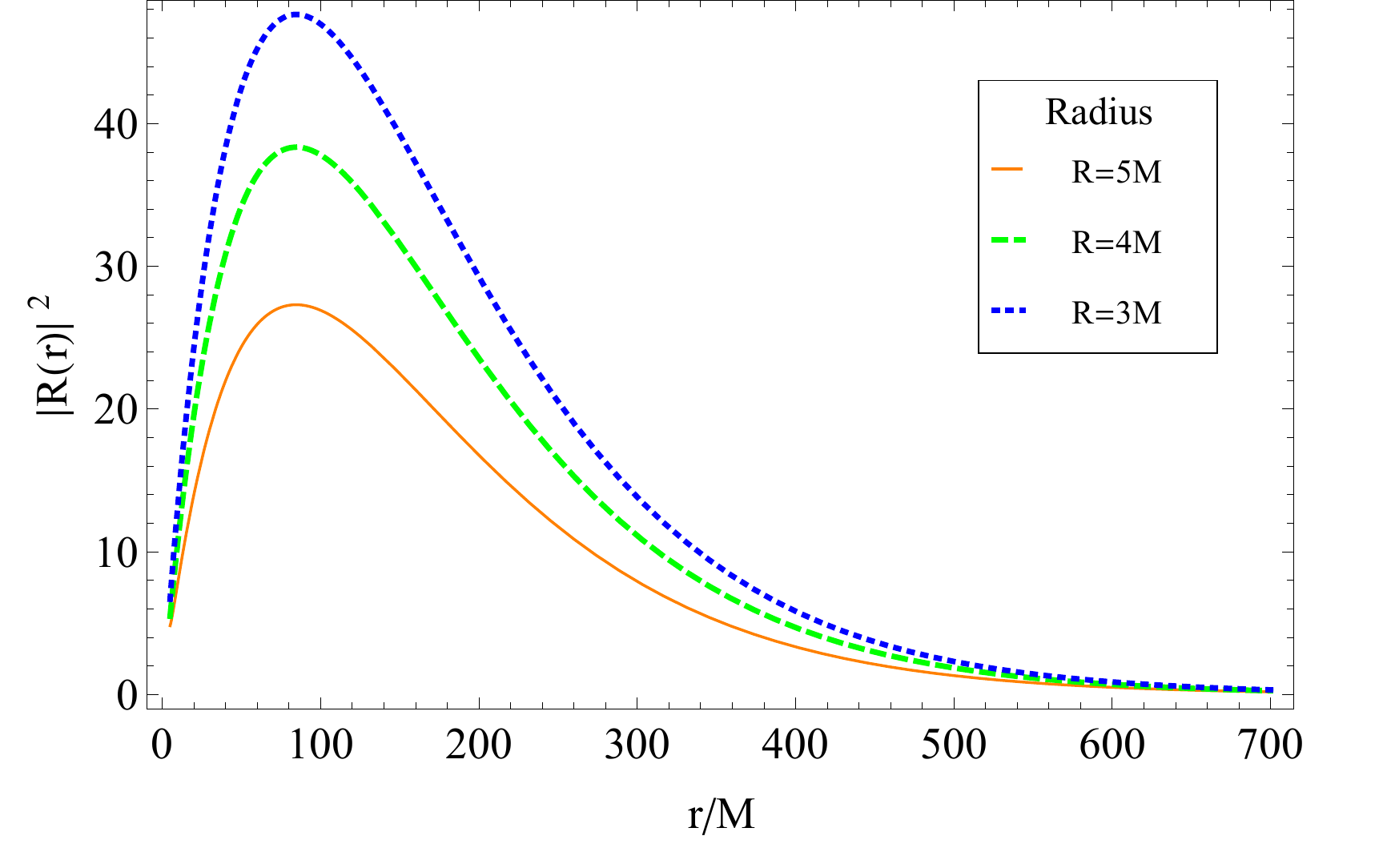}
\caption{Quasi Boundstates for $\mu M=0.15$, $\alpha M=5$, $\Omega M=0.1$, $l=m=1$.}
\label{starqbsR}
\end{minipage}
\ \ \ \ \
\begin{minipage}{.5\textwidth}
\centering
\includegraphics[width=8cm]{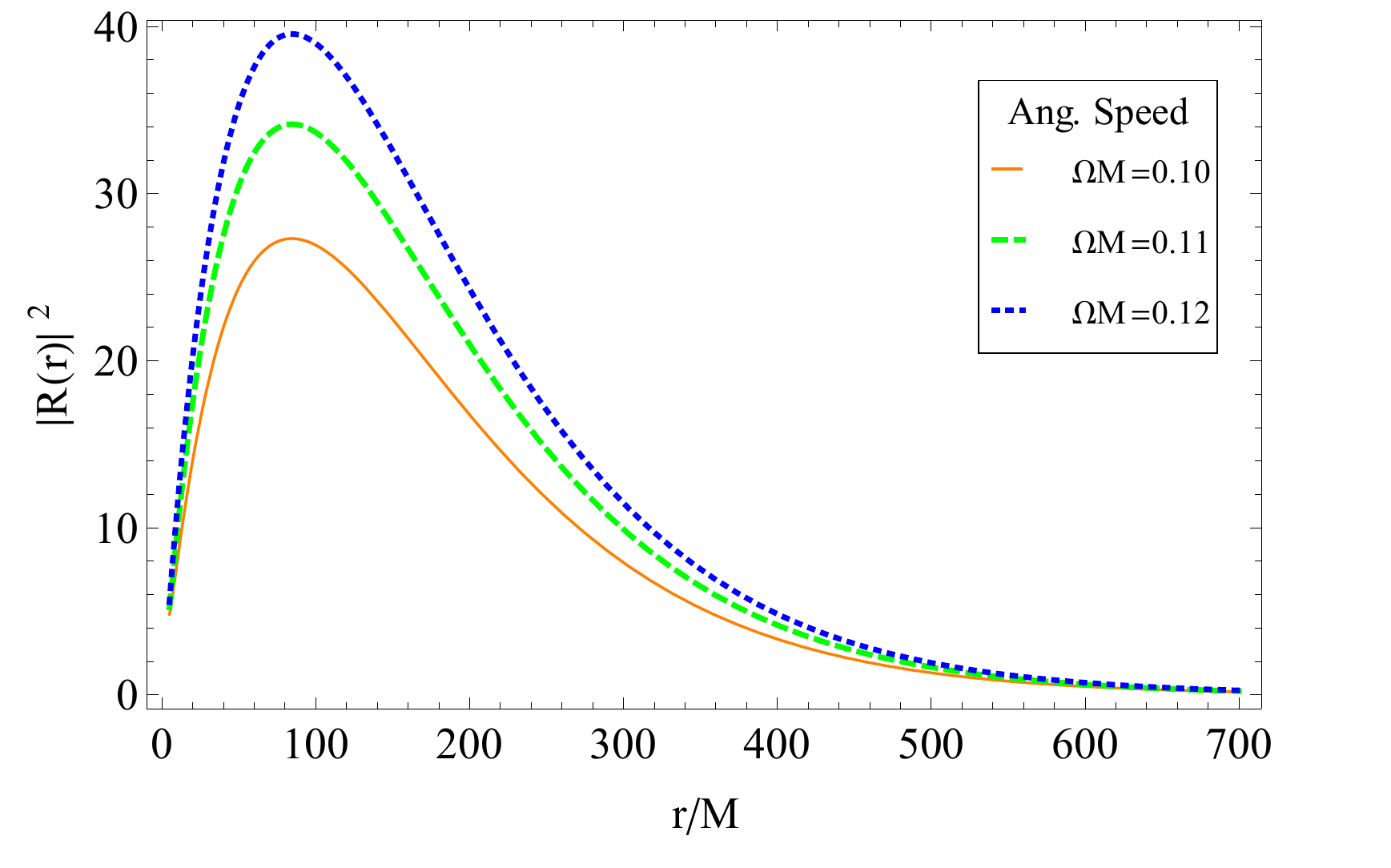}
\caption{Quasi Boundstates for $\mu M=0.15$, $\alpha M=5$, $R=5M$, $l=m=1$.}
\label{starqbsO}
\end{minipage}
\end{figure}

\begin{figure}[h!]
\begin{minipage}{.5\textwidth}
\centering
\includegraphics[width=8cm]{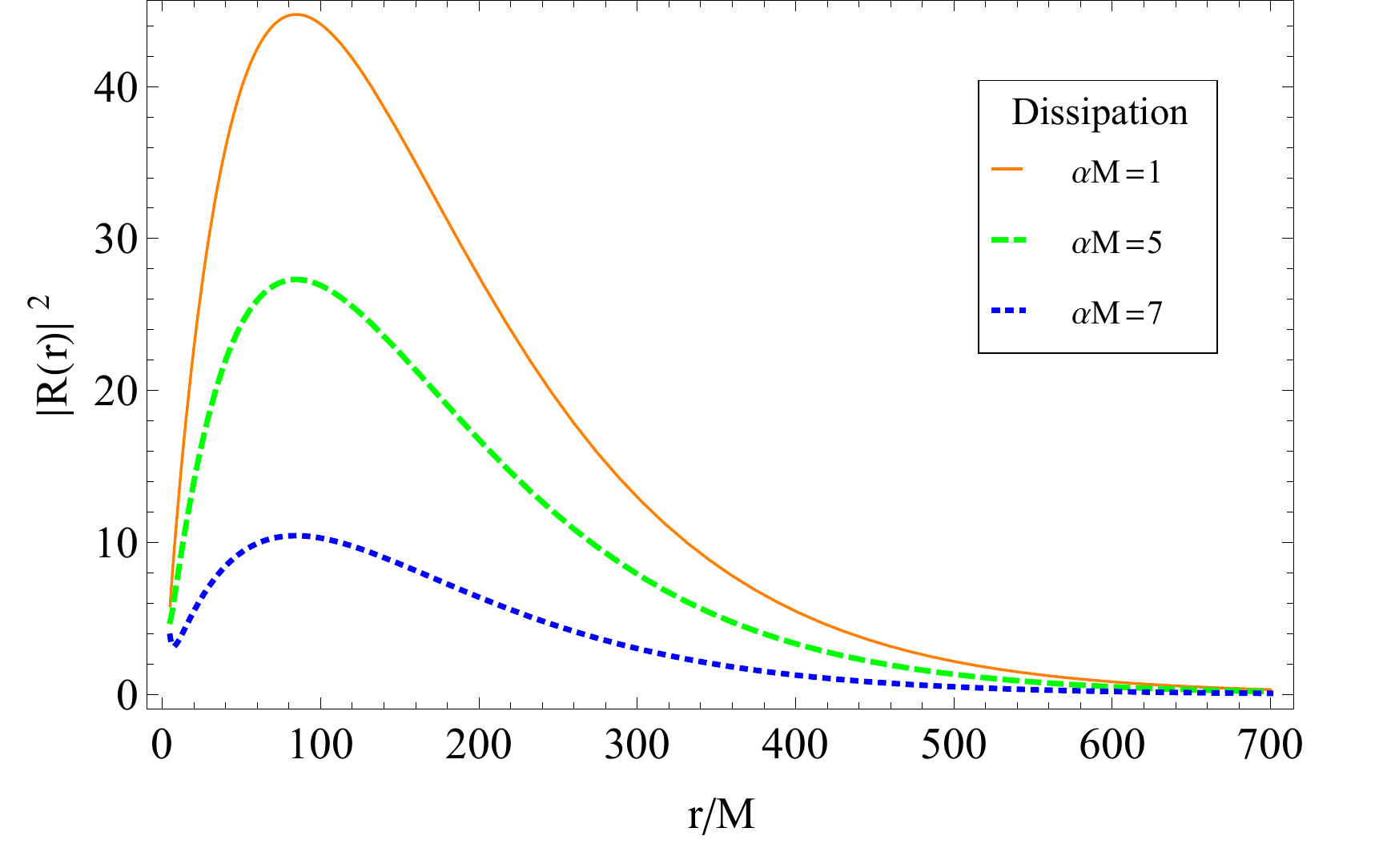}
\caption{Quasi Boundstates for $\mu M=0.15$, $\Omega M=0.1$, $R=5M$, $l=m=1$.}
\label{starqbsA}
\end{minipage}
\ \ \ \ \
\begin{minipage}{.5\textwidth}
\centering
\includegraphics[width=8cm]{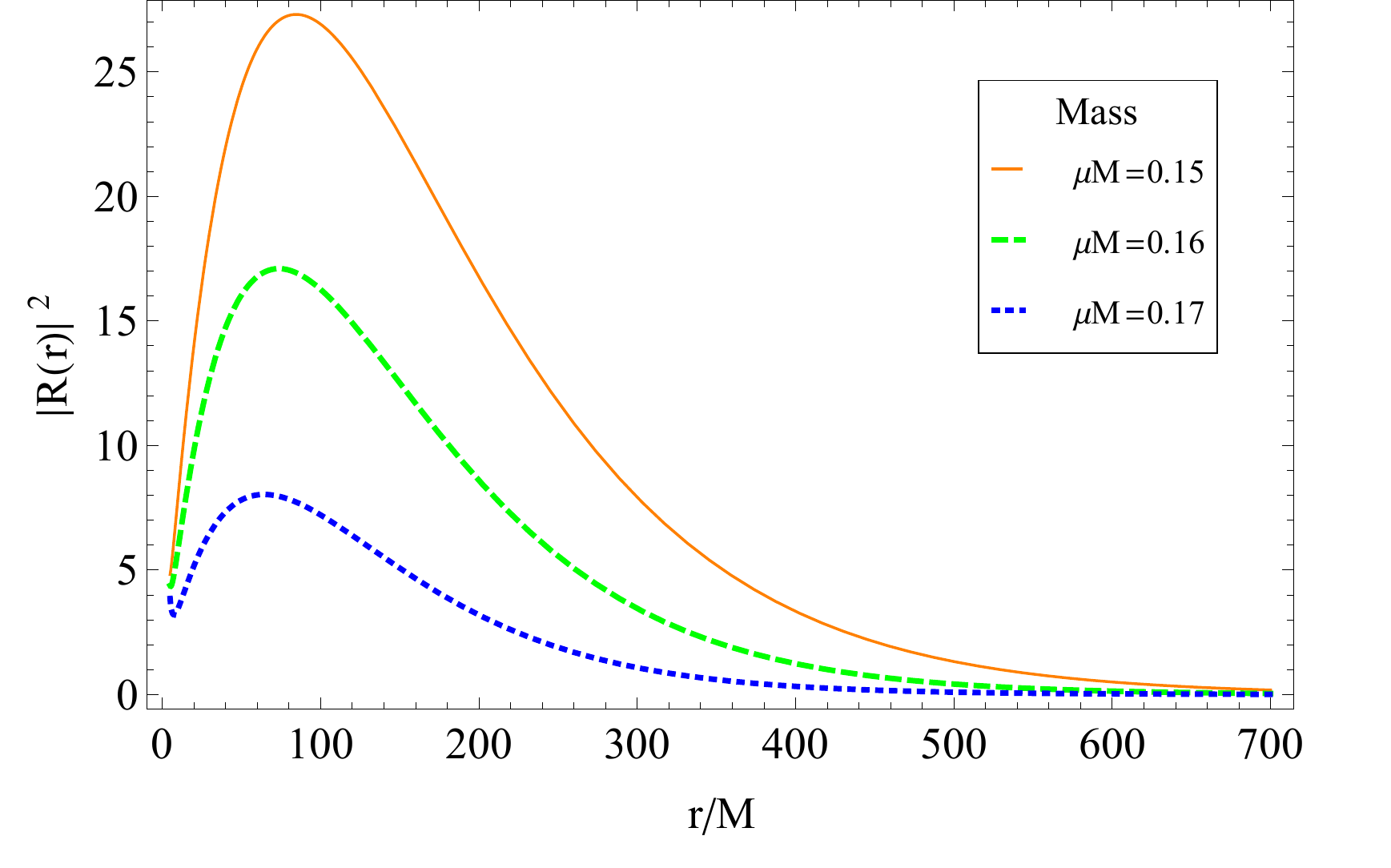}
\caption{Quasi Boundstates for $\Omega M=0.1$, $\alpha M=5$, $R=5M$, $l=m=1$.}
\label{starqbsM}
\end{minipage}
\end{figure}

The quasi-boundstates were also obtained for different values of the star radius, angular momentum, dissipation and perturbation mass. Unlike the previous BH case, we do not have an analytical value to use as a first guess for the real part of the frequency, which makes it more difficult to find the states. However, these states were found by performing slight changes on the value of the frequency to verify if the wavefunction becomes more or less stable after the change, and by reppeating this process up until it was possible to find a guess close enough to the correct value for the numerical calculations to find the root. These results are shown in figures \ref{starqbsR}, \ref{starqbsO}, \ref{starqbsA}, and \ref{starqbsM}.

Just like in the previous chapter, we show that an increase in the angular momentum does not change the radial position of the QBS peak, but an increase in the mass $\mu$ compresses the QBS to smaller values of $r$. Also, we show that an increase in the dissipation and star radius do not alter the radial position of the peak, which is somehow expected since we verified that an increase in one of these variables leads to an increase of the superradiance, and therefore their effect on the QBS should be similar to the effect of $\Omega$.

\begin{figure}[h!]
\begin{minipage}{.5\textwidth}
\centering
\includegraphics[width=8cm]{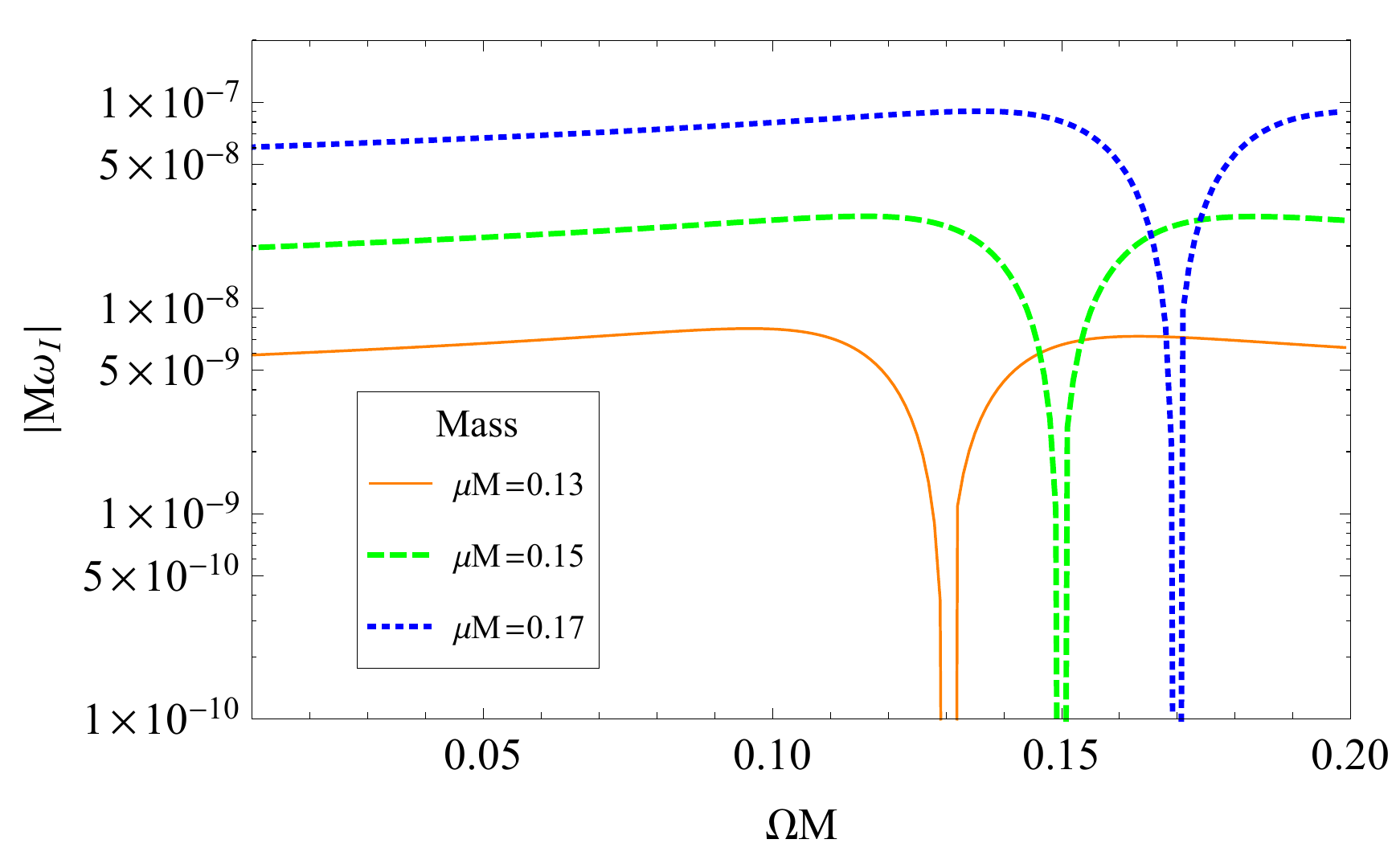}
\caption{Absolute value of the imaginary part of the QBS frequency for $\alpha M=20$, $R=4M$, $l=m=1$.}
\label{starinstamu}
\end{minipage}
\ \ \ \ \
\begin{minipage}{.5\textwidth}
\centering
\includegraphics[width=8cm]{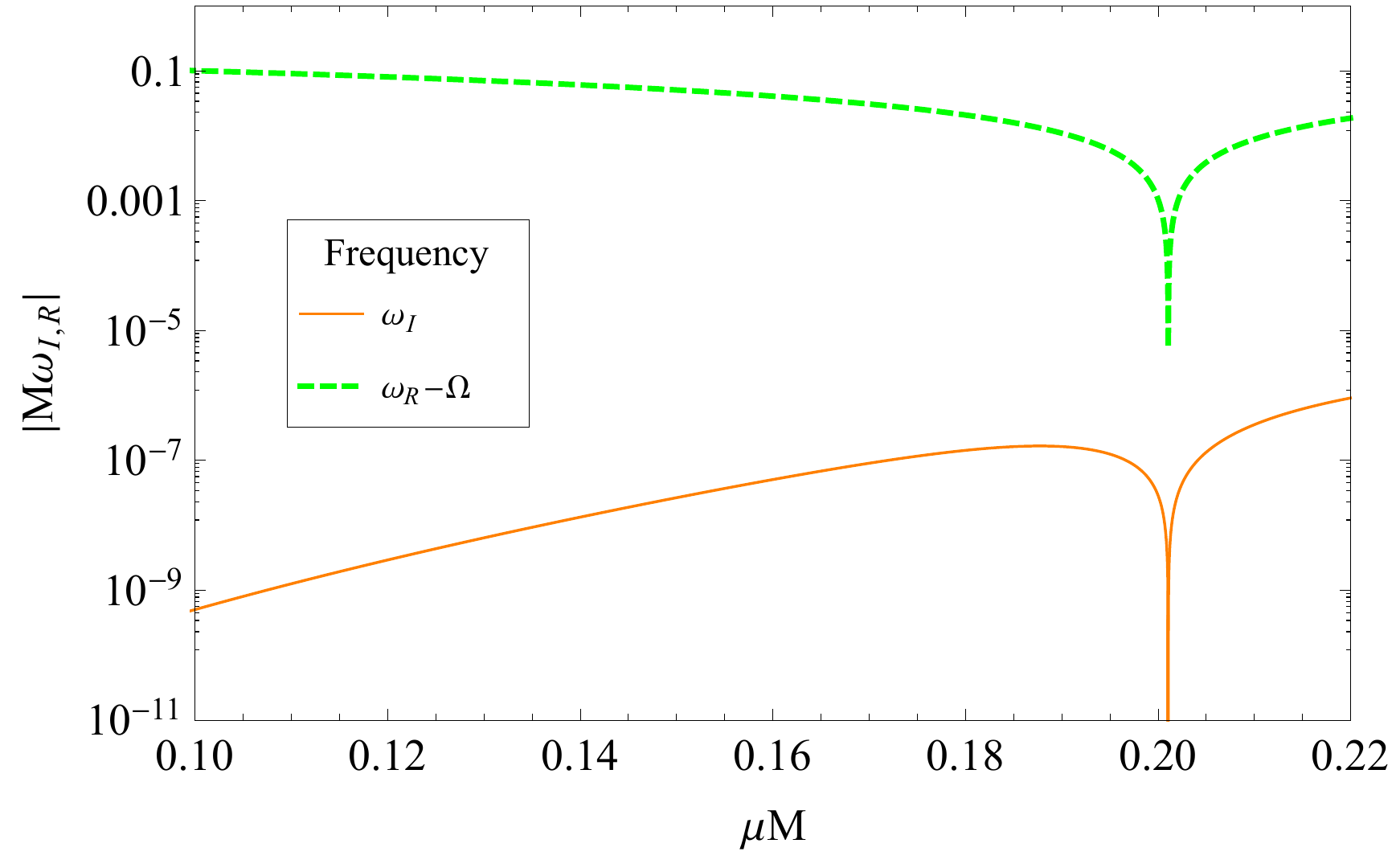}
\caption{Absolute value of the imaginary part of the QBS frequency for $\alpha M=20$, $R=4M$, $\Omega M=0.2$, $l=m=1$.}
\label{starinsta}
\end{minipage}
\end{figure}

The imaginary part of the frequency was computed as a function of $\Omega$ for different field masses. This result is shown in figure \ref{starinstamu}. Like in the BH case, we verify that there is a value, which increases with the field mass, for which $\omega_I$ changes its sign. We also plotted $\omega_I$ and $\omega_R-m\Omega$ as functions of $\mu$ to show, like in the BH case, that the turning point of the sign of $\omega_I$ corresponds to the point where $\omega_R=m\Omega$, which is the threshold frequency for the superradiant region. These results are shown in figure \ref{starinsta}.

\begin{figure}[h!]
\begin{minipage}{.5\textwidth}
\centering
\includegraphics[width=8cm]{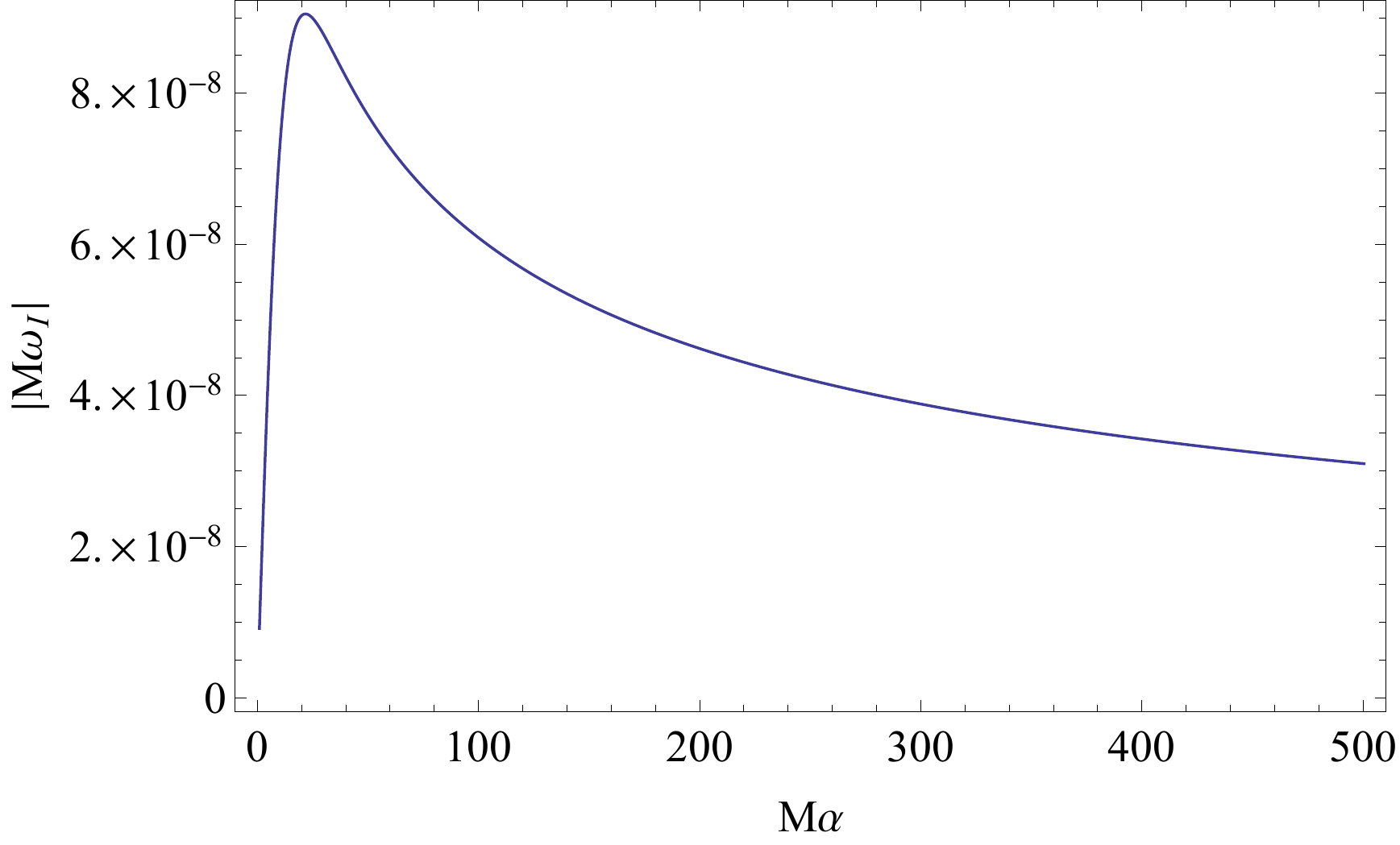}
\caption{Absolute value of the imaginary part of the QBS frequency for $\alpha M=20$, $\mu M=0.15$, $\Omega M=0.2$, $l=m=1$.}
\label{superinsta}
\end{minipage}
\ \ \ \ \
\begin{minipage}{.5\textwidth}
\centering
\includegraphics[width=8cm]{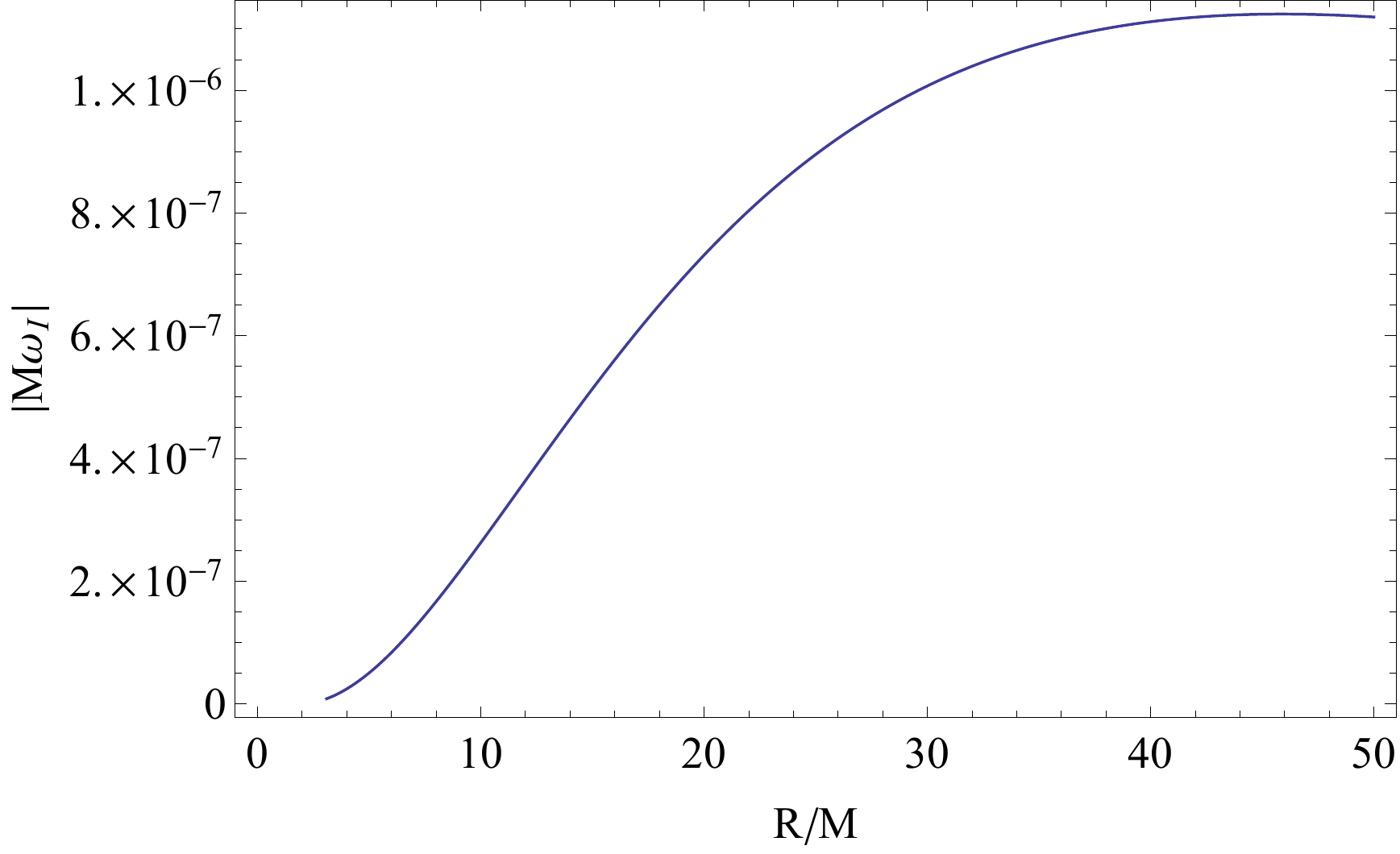}
\caption{Absolute value of the imaginary part of the QBS frequency for $\mu M=0.15$, $R=4M$, $\Omega M=0.2$, $l=m=1$.}
\label{starinstR}
\end{minipage}
\end{figure}

Finaly, we show a plot of $\omega_I$ as a function of the radius $R$ and the dissipation coefficient $\alpha$, as can be seen in figures \ref{superinsta} and \ref{starinstR}. Our results seem to indicate that there is a value of $\alpha$ that maximizes the instability. We also show that the instability increases with the radius of the star, but it seems to stagnate at around $R=50M$ and $R=3M$.

\subsection{Comparison with analytical results}

\begin{figure}[h!]
\centering
\includegraphics[width=9cm]{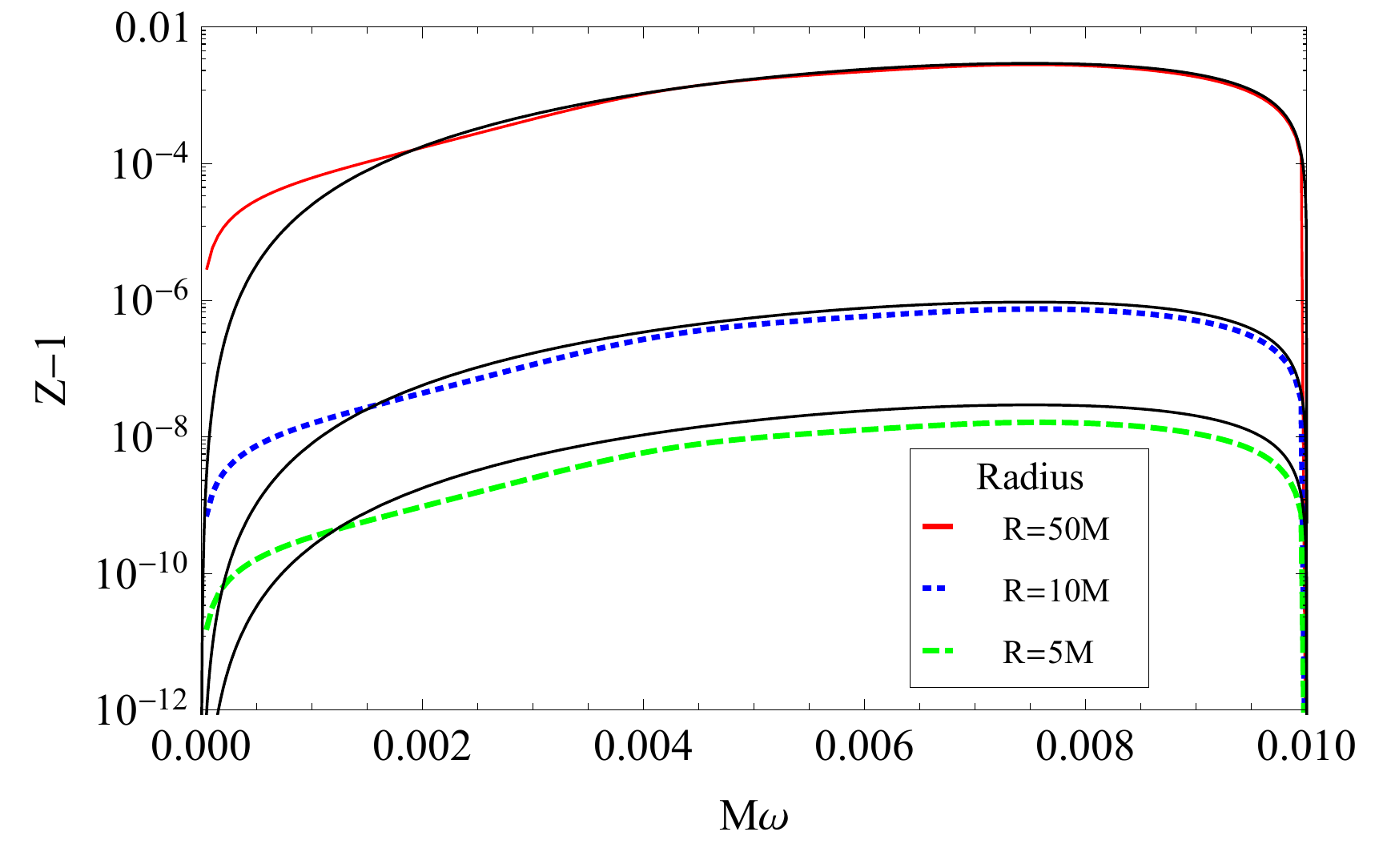}
\caption{Comparison of the reflection coefficients obtained numerically with the analytical results.}
\label{analytical}
\end{figure}

For non-relativistic Newtonian configurations, we can take the limits $M/R<<1$ and $\Omega R<<1$ in order to obtain an equation which can be solved analytically inside and outside the star:
\be
{R}''\left(r\right)+\frac{2}{r}{R}'\left(r\right)+\left(\omega^2-i\omega\alpha-\frac{\lambda}{r^2}\right)R\left(r\right)=0.
\ee
From this point, noting that $\lambda=l\left(l+1\right)$, multiplying through by $r^2$ and performing a coordinate change of the form ${r}'=r\sqrt{\omega^2-i\omega\alpha}$, we obtain the spherical bessel equation given by (dropping the primes):
\be
r^2{R}''\left(r\right)+2r{R}'\left(r\right)+\left[r^2-l\left(l+1\right)\right]R\left(r\right)=0.
\ee
The solution of this equation is a linear combination of the spherical Bessel functions $y_n\left(x\right)$ and $j_n\left(x\right)$. Outside the star, both terms of this combination exist because they are regular throughout the entire range of the radial coordinate. However, the function $y_n\left(x\right)$ diverges at the origin, and therefore this term must vanish inside the star, which leaves the coefficient of $j_n$ arbitrary, that we set to be $1$. We are left with:
\be
R_{int}\left(r\right)=j_l\left(r\sqrt{\left(\omega-m\Omega\right)\left(i\alpha+\omega-m\Omega\right)}\right),
\ee
\be
R_{ext}\left(r\right)=A j_l\left(r\omega\right) + B y_l\left(r\omega\right).
\ee
Since we are intrested in computing the amplification factors, we must write our wave function outside the star as a linear combination of ingoing and outgoing waves. The linear combination of spherical Bessel functions can be written in more convenient way:
\be
R_{ext}\left(r\right)=\frac{i^{-l}\left(A-iB\right)\left(l+l^2-2ir\omega\right)}{4r^2\omega^2}e^{ir\omega}+\frac{i^l\left(A+iB\right)\left(l+l^2+2ir\omega\right)}{4r^2\omega^2}e^{-ir\omega}=A_{\infty}e^{ir\omega}+B_{\infty}e^{-ir\omega}.
\ee
The values of A and B can be obtained by performing a matching of the interior and exterior solutions and their derivatives at the surface of the star, which should be continuous. Then, computing the amplification factor of the form \eqref{amp} we obtain, at lowest order in $R\Omega$:
\be
Z=1+\frac{4\alpha R^2\left(\Omega-\omega\right)\left(\omega R\right)^{2l+1}}{\left(2l+1\right)!!\left(2l+3\right)!!}.
\ee
This result is of extreme importance since it allows us to verify the validity of our results. For example, for $l=1$ we verify that our numerical results agree extremely well with the analytical ones (black lines), with a small difference in the low frequency limit, as can be seen in figure \ref{analytical}. This seems to indicate that the mass of the star is only important in the low frequency region.
Also, considering the same $l=1$ case, it is possible to recover the analytical result for slowly rotating black holes by imposing $R=2M$ and, since it is known that the units of $\alpha$ are the inverse of a distance, $\alpha=1/M$, as it is the only possible timescale at this point:
\be
Z=1+\frac{16}{45}M\left(\Omega-\omega\right)\left(2M\omega\right)^3,
\ee
which is very similar to the full GR calculation, with just a slight difference on the numerical coefficient (that is $2/9$)~\cite{starobinski}.

\section{Relativistic perfect-fluid star}

To study the relativistic effects of a rotating star we shall add a non diagonal term to the metric tensor $g_{t\phi}$. The metric tensor that describes a perfect-fluid rotating star in GR is given by~\cite{hartle}:
\be
ds^2=-e^{2\varphi}dt^2+\left(1-\frac{2m\left(r\right)}{r}\right)^{-1}dr^2+r^2\left[d\theta^2+\sin^2\theta \left(d\phi-\zeta\left(r\right)dt\right)^2\right],
\ee
where $\zeta\left(r\right)$ is the angular speed of an inertial frame at a distance $r$ of the center of the star as seen from an observer at the infinity. To simplify the calculations, it is assumed that this angular speed is small and only first order terms are considered. To deduce a differential equation to describe $\zeta\left(r\right)$, only the following field equation is required:
\be\label{field}
R_\phi^t=8\pi T_\phi^t.
\ee

In this case the stress-energy tensor \eqref{stress} is slightly different. The quadrivelocity vector has a new entry $U^\phi=\Omega U^t$. Leaving $U^\phi=\Omega U^t$ and using $U_aU^a=-1$ we can compute the entry $U^t$ which is:
\be
U^t=\left[-\left(g_{tt}+2\Omega g_{t\phi}+\Omega^2g_{\phi\phi}\right)\right]^{-\frac{1}{2}}.
\ee

\subsection{Frame-dragging}

Now, to obtain an equation for $\zeta\left(r\right)$, lets define $g_{tt}=f\left(r\right)$ and $g_{rr}=B\left(r\right)$. Expanding the field equation \eqref{field}, dividing through by $-\frac{r^2\sin^2\theta}{2B\left(r\right)f\left(r\right)}$ and keeping only first order terms in $\zeta\left(r\right)$, we obtain:
\be
{\zeta}''\left(r\right)+\frac{1}{2}\left(\frac{8}{r}-\frac{{B}'\left(r\right)}{B\left(r\right)}-\frac{{f}'\left(r\right)}{f\left(r\right)}\right){\zeta}'\left(r\right)=16\pi\left(\rho+P\right)\left(\zeta\left(r\right)-\Omega\right)B\left(r\right).
\ee
Outside the star, the metric reduces to the Schwarzschild metric and the stress-energy tensor vanishes. This simplifies the previous equation to
\be
{\zeta}''\left(r\right)+\frac{4}{r}{\zeta}'\left(r\right)=0,
\ee
which can be solved exactly to obtain
\be
\zeta_{out}\left(r\right)=\frac{2J}{r^3},
\ee
where $J$ is a constant of integration that can be identified to be the total angular momentum of the star ~\cite{papapetrou1948}. Inside the star, this equation can not be solved analytically due to its non-trivial dependence in $r$. Therefore, we perform a variable transformation of the form $\bar\zeta\left(r\right)=\Omega-\zeta_{int}\left(r\right)$ and the equation is solved numerically.

The same numerical problem met in the previous sections is present, since we still can not start our integration from $r=0$. Therefore, we perform a series expansion for $\bar\zeta\left(r\right)$ around the origin and use that series as a boundary condition to start the integration. The expansion is given by:
\be
\zeta_0\left(r\right)=\sum_n Z_n r^{2n},
\ee
where we are considering only even powers of $r$ in this expansion because it was verified that the coefficients $Z_n$ for odd $n$ are identically equal to zero. The coefficient $Z_0$ can be computed since our equation imposes that $\bar\zeta_0\left(0\right)=\zeta_c$ is a constant.

It is now possible to obtain a function $\zeta\left(r\right)$ that holds for the entire space. In practice, the integration is started outwards from the center of the star where we impose as boundary conditions $\bar\zeta\left(0\right)=\zeta_c$ and ${\bar\zeta}'\left(0\right)=0$. When the surface of the star is reached, it is possible to compute the total angular momentum $J$ and angular speed $\Omega$ by matching $\bar\zeta\left(R\right)$ with $\zeta_{out}\left(R\right)$, that is:
\be
J=\frac{1}{6}R^4{\bar\zeta}'\left(R\right),\ \ \ \ \ \Omega=\bar\zeta\left(R\right)+\frac{2J}{R^3}.
\ee
Now that $\Omega$ is known, one can perform a coordinate transformation for a co-rotating frame like before and solve the wave equation.

\subsection{The wave equation}

Once again, inserting the ansatz \eqref{ansatz} into the modified Klein-Gordon equation, expanding and separating the sum inside the derivative, and substituting the inverse of the metric, the exponential terms cancel out and we obtain an equation which is separable in $r$ and $\theta$. Dividing through by $R\left(r\right)S\left(\theta\right)e^{2\varphi}$ we obtain the equations:
\be
f\left(r\right)\left[e^{-\varphi}{e^{\varphi}}'+\frac{2}{r}+f^{-1}\left(r\right)\left(\frac{m\left(r\right)}{r^2}-\frac{{m}'\left(r\right)}{r}\right)\right]\frac{r^2}{R}\partial_rR+e^{-2\varphi}r^2\omega\left(\omega+2m\zeta\left(r\right)\right)+f\left(r\right)\frac{r^2}{R}\partial_r\partial_rR-\mu^2r^2-i\omega\alpha r^2=\lambda;\nonumber 
\ee
\be
f\left(r\right)=\left(1-\frac{2m\left(r\right)}{r}\right),
\ee
for the radial equation, and:
\be
-\cot\theta\frac{\partial_\theta S}{S}+\frac{m^2}{\sin^2\theta}-\frac{\partial_\theta\partial_\theta S}{S}=\lambda,
\ee
for the angular equation. The angular equation can be solved exactly by the spherical harmonics and therefore we shall not study it again. The radial equation can be multiplied through by $\frac{R}{r\left(r-2m\left(r\right)\right)}$ in order to obtain a new radial equation of the form \eqref{rad}, where the coefficients are given by: 
\be
A_r\left(r\right)=e^{-\varphi}{e^{\varphi}}'+\frac{2}{r}+\left(1-\frac{2m\left(r\right)}{r}\right)^{-1}\left(\frac{m\left(r\right)}{r^2}-\frac{{m}'\left(r\right)}{r}\right),
\ee
\be
B_r\left(r\right)=\left(1-\frac{2m\left(r\right)}{r}\right)^{-1}\left(\omega\left(\omega+2m\zeta\left(r\right)\right)e^{-2\varphi}-\mu^2-i\omega\alpha-\frac{\lambda}{r^2}\right).
\ee
Similarly to the previous case, we still need to perform a transformation $\omega\to\omega-m\Omega$.

The outside metric is no longer a Schwarzschild metric. However, since outside the star the angular speed of the frame dragging decreases with $r^{-3}$, this term is neglectable at infinity and the series expansion is the same as before (see \eqref{inf1},\eqref{inf2}). At the origin, we know that the angular speed of the frame dragging converges to a finite value $\zeta_c$, and therefore in the limit $r\to 0$ this term is neglectable against the $\lambda$ one. For this reason, the series expansion at the origin is also the same as before (see \eqref{ori}). The coefficients of these series are again an arbitrary $C_0$ that we set to be 1, and $A_\infty$ and $B_\infty$ are unknown.

\subsection{Numerical results}

The procedure to solve the radial equation numerically is exactly the same as before. However, this time we must compute the angular speed $\Omega$ and the angular speed of the frame-dragging effect $\zeta\left(r\right)$ before doing so, since the coefficients of the differential equation depend on these variables. 

\begin{figure}[h!]
\centering
\includegraphics[width=9cm]{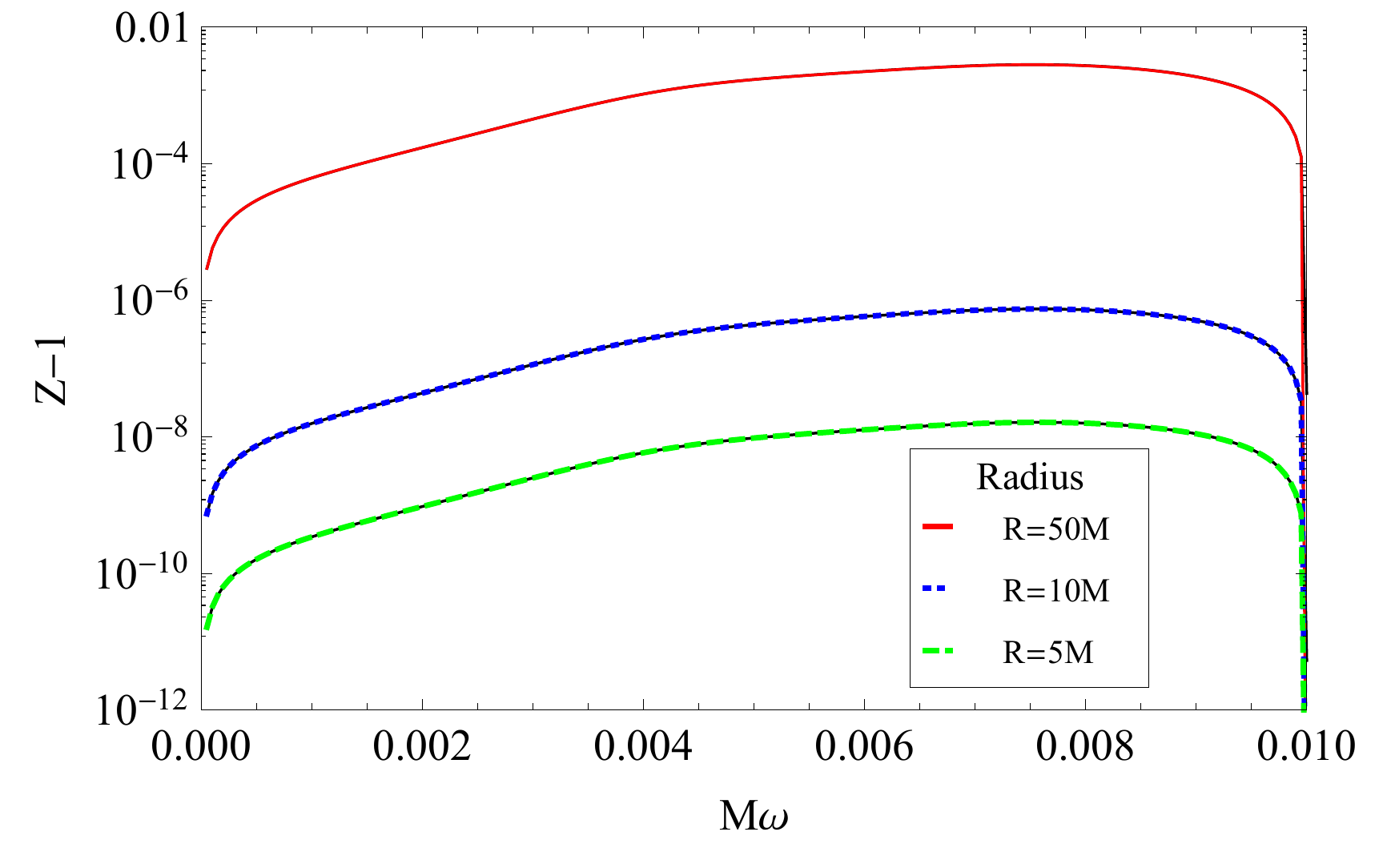}
\caption{Comparison of the reflection coefficients obtained numerically with the relativistic results}
\label{relativistic}
\end{figure}

The amplification factors ~\eqref{amp} were obtained for the same parameter values as in the previous section, and are shown in figure \ref{relativistic}. We can see that the results are very similar to the previous ones. This is an indication that the relativistic effects related to frame-dragging are not relevant in the calculation of the amplification factors and the study of the phenomenon of superradiance. This is an indication that the most important phenomenon responsible for the existence of superradiance is dissipation.

\cleardoublepage





\chapter{Conclusions}
\label{chapter:conclusions}

In this work it was shown that rotational superradiance is a process that can be studied in the context of General Relativity and, in particular, black-hole physics and astrophysics. It was shown that this process is related to rotating systems, like the Kerr BH, and also to dissipative phenomena, since rotating stars are shown to display superadiance when dissipation is properly included. 

From equation \eqref{staby} we can use our results to state that there are no unstable modes for the Schwarzschild BH, because at $a=0$ the value of $\omega_I$ is negative for any field mass, and thus the scalar field decreases exponentially with time. However, for the Kerr BH, $\omega_I$ is only negative in certain regions of low angular momentum and large field mass. As we decrease the mass of the field, a BH with a certain angular momentum becomes progressively less stable and, when $\omega_I$ changes its sign, it becomes unstable. The same behaviour is obtained if we fix the mass of the field and increase the angular momentum instead. The main conclusion is that the Kerr BH stability decreases as the perturbation mass decreases (this is valid in the $M\mu<<1$ limit) and the angular momentum increases. It was also verified that the value of $a$ at which $\omega_I$ changes its sign is given by the threshold frequency of equation \eqref{super_condition}, which means that these instabilities only occur in the superradiant regime, as expected. 

Also from equation \eqref{staby}, we can state that there are no unstable modes for non-rotating stars, because in the limit $\Omega=0$, where ${\omega}'=\omega$, the imaginary part of the frequency $\omega_I$ is always negative. Also, for a rotating stable star with a given angular momentum, as we decrease the mass of the perturbation, it becomes progressively less stable until $\omega_I$ changes its sign, and it becomes unstable. It was also verified once again that the value of $\omega_R$ at which $\omega_I$ changes its sign is given by the threshold frequency \eqref{super_condition}, which tells us that these instabilities only occur in the superradiant regime.

For $\mu=0$, we have found that the imaginary part of the frequency is very close to zero. However, since the QBSs are formed for frequencies $\omega<\mu$, at the limit $\mu=0$ the real part of the frequency is also $\omega_R=0$, and these states do not exist. This is not a complete proof of the stability of the Kerr BH since in this case we must study the quasi-normal modes in order to find stable or unstable solutions, but it tells us that there are no unstable boundstates and therefore the solution should be stable.

Finally, the study of BH superradiance was proven to be a very rich and extremely useful phenomenology. The next step of this study was to verify if these processes might also take place in other self-gravitating systems such as stars, and the study was a success: stars do display superradiant phenomena with implications on their stability. This work opens the door to probe new physics beyond the standard model, constraint the interaction cross-section between dark matter ultra-light degrees of freedom with ordinary matter, or the self-annihilation cross-section. These results also make it possible the existence of stationary configurations describing a star covered by a scalar condensate and solely supported by superradiance. It was also shown that this phenomenon is present even in non-relativistic systems, and that the relativistic effects related to frame-dragging are neglectable in this study.

\cleardoublepage




%
%

\cleardoublepage

\cleardoublepage

\appendix

\chapter{The Klein-Gordon equation}
\label{anexo}

The Einstein's vacuum equations can be obtained by applying the variational method to the Einstein-Hilbert action. To study a perturbation, one must couple this action to a field that describes it before obtaining the equations of motion. This perturbation, which consists of a small deviation of the background geometry, is also described by an equation of motion and the solution of this equation dictates the stability of the system against that perturbation. Notice that the perturbation must be small because otherwise we would no longer be working in the same initial background geometry. In this chapter, the equation of motion for a massive complex scalar field is deduced. 

Since we are studying a scalar perturbation, we shall minimally couple the action to a massive scalar field $\Psi$:
\be
S=\int_\Omega\mathcal{L}d\Omega=\int_\Omega \sqrt{-g}\left[\frac{R}{\kappa}-\frac{1}{2}g^{ab}\partial_a\Psi^*\partial_b\Psi-\frac{\mu^2}{2}\Psi^*\Psi\right]d\Omega,
\ee
where $R$ is the Ricci's scalar, $\kappa$ is the coupling constant, $\mu^2$ is the mass of the scalar field, $g_{ab}$ is the metric tensor, and $g$ is the metric determinant.

The equations of motion for a given field are obtained by minimizing the action, that is, differentiating the action in order to the field in study and equaling the result to zero:
\be
\label{variat}
\delta S=\int_\Omega\left(\frac{\partial\mathcal{L}}{\partial\Psi}\delta\Psi\right) d\Omega = 0.
\ee

We shall differentiate in order to $\Psi^*$ to obtain directly an equation of motion for $\Psi$. Applying the previous method to the action given we obtain
\be
\int_\Omega \sqrt{-g} \left(-\frac{1}{2}g^{ab}\partial_b \Psi \partial_a \delta\Psi^* -\frac{\mu^2}{2}\Psi\delta\Psi^*\right)d\Omega=0.
\ee

The second term in the integral is already written in the form \eqref{variat}, but the first term still deppends on the derivative of our field. Integrating that term by parts we obtain
\be
\int_\Omega \left[ \frac{1}{2} \partial_a \left(\sqrt{-g} g^{ab} \partial_b \Psi\right) - \sqrt{-g} \frac{\mu^2}{2}\Psi\right] \delta\Psi^* d\Omega = 0,
\ee
since the integral at the boundary of $\Omega$, the infinity, is equal to zero. This happens because we will be working with asymptotically flat solutions of the Einstein field equations. Since this equation must hold for an arbitrary scalar field $\Psi^*$, the function inside the integral must be identically zero from which we obtain
\be
\frac{1}{\sqrt{-g}}\partial_a \left(\sqrt{-g} g^{ab} \partial_b \Psi\right)=\mu^2\Psi.
\ee

Now, to write the left side of the previous result in a covariant description, note that $\Psi$ is a scalar field, and therefore its partial derivative is equal to the covariant derivative. Furthermore, using the fact that $\partial_a\sqrt{-g}=\Gamma^b_{ba}\sqrt{-g}$, the product derivative of this term can be written in the form
\be
\frac{1}{\sqrt{-g}}\partial_a \left(\sqrt{-g} g^{ab} \nabla_b \Psi\right)=\Gamma^b_{ba}\nabla^a\Psi+\partial_a\nabla^a\Psi,
\ee
and using the definition for the covariant derivative of a rank-1 covariant tensor we obtain 
\be
\nabla_a\nabla^a\Psi=\mu^2\Psi,
\ee
which is the Klein-Gordon equation for the scalar field $\Psi$ that we were trying to obtain. Assuming that our metric tensor can be written in the following general form:
\be
ds^2=-g_{tt}dt^2+g_{rr}dr^2+g_{\theta\theta}d\theta^2+g_{\phi\phi}d\phi^2+2g_{t\phi}dtd\phi,
\ee
it is possible to expand the Klein-Gordon equation with the ansatz \eqref{ansatz}. It is straightforward to see that we can write the equation as:
\beq
S\left(\theta\right)R\left(r\right)\left(\partial_t \partial^t+\partial_\phi \partial^\phi\right) \left( e^{-im\phi-i\omega t}\right)+
S\left(\theta\right) e^{-im\phi-i\omega t} \partial_r \partial^r R\left(r\right) + 
R\left(r\right) e^{-im\phi-i\omega t} \partial_\theta \partial^\theta S +\\ \nonumber
+ \frac{1}{\sqrt{-g}}\left( S\left(\theta\right) e^{-im\phi-i\omega t} \partial^r R\left(r\right) \partial_r \sqrt{-g}+
  R\left(r\right) e^{-im\phi-i\omega t} \partial^\theta S\left(\theta\right) \partial_\theta \sqrt{-g}\right)=
\mu^2 S\left(\theta\right)R\left(r\right)e^{-im\phi-i\omega t}.
\eeq
Computing the derivatives in order to $t$ and $\phi$, the exponential factor cancels everywhere. Dividing through by $R\left(r\right)S\left(\theta\right)$ we obtain a simpler expression given by
\beq
&&-\omega^2 g^{tt}-
2m\omega g^{t\phi}-
m^2 g^{\phi\phi}+
\partial_r g^{rr} \frac{\partial_r R\left(r\right)}{R\left(r\right)}+
g^{rr} \frac{\partial_r \partial_r R\left(r\right)}{R\left(r\right)}+
\partial_\theta g^{\theta\theta}\frac{\partial_\theta S\left(\theta\right)}{S\left(\theta\right)}+\\ \nonumber
&&+g^{\theta\theta}\frac{\partial_\theta \partial_\theta S\left(\theta\right)}{S\left(\theta\right)}+
g^{rr}\frac{\partial_r R\left(r\right)}{R\left(r\right)}\frac{\partial_r\sqrt{-g}}{\sqrt{-g}}+
g^{\theta\theta}\frac{\partial_\theta S\left(\theta\right)}{S\left(\theta\right)}\frac{\partial_\theta\sqrt{-g}}{\sqrt{-g}}=
\mu^2.
\eeq

Both systems in study in this thesis can be described by metric tensors in the form used to deduce this last equation. However, this is the furthest we can go without defining the entries of the metric tensor. In the case of the slowly-rotating perfect fluid star, this equation gets even simpler because $g_{t\phi}=0$.


\cleardoublepage



\chapter{Pressure of a perfect fluid}

In the study of superradiance in stars we made use of the stress-energy tensor that describes a perfect fluid, given by equation \eqref{stress}, where $\rho$ and $P$ are the density and the pressure of the fluid, respectively, and $U^a$ is the quadrivelocity vector, given by $U^t=\sqrt{-g_{tt}}$ and $U^r=U^\theta=U^\phi=0$. Assuming that the density of the fluid is constant and that for $r>R$, where $R$ is the radius of the star, the metric tensor is the Schwarzschild solution, we can easily obtain
\be
M=\frac{4}{3}\pi R^3 \rho\ \Leftrightarrow\ \rho=\frac{3M}{4\pi R^3}.
\ee
However, to obtain the pressure as a function of $\rho$ and $r$, we have to compute the Einstein's tensor $G_{ab}$ and solve a field equation that we choose to be
\be\label{field}
G_{rr}=8\pi T_{rr}.
\ee
If we assume that the metric tensor can be written in the general static and spherically symmetric form
\be
ds^2=-A\left(r\right)^2dt^2+B\left(r\right)^2dr^2+r^2\left(d\theta^2+\sin^2\theta d\phi^2\right),
\ee
a long but straightforward calculation (it can be done either by computing the Christoffel symbols or using Cartan's structure equations) leads us to 
\be
G_{rr}=\frac{2}{rAB^2}\frac{dA}{dr}-\frac{B^2-1}{r^2B^2}.
\ee
Now, if we replace the metric tensor given by \eqref{fluid} in the previous result and solve the field equation \eqref{field}, we obtain
\be
P=\rho\left(\frac{\sqrt{1-2Mr^2/R^3}-\sqrt{1-2M/R}}{3\sqrt{1-2M/R}-\sqrt{1-2Mr^2/R^3}}\right).
\ee

\cleardoublepage


\chapter{Viscous gravity analogue}

A different way to include dissipation in a gravitational system is to use the methods of analogue gravity. For example, linear perturbations in an inviscid, irrotational flow propagate as fields on a curved space-time. Thus, one natural option to model dissipative phenomena is through the inclusion of a kinematic viscosity. For a fluid at rest, the wave equation for the perturbations is given by~\cite{liberati}
\be\label{visc}
\partial_t^2\phi=\nabla^2\phi+\frac{4}{3}\nu\partial_t\nabla^2\phi.
\ee

\begin{figure}[h!]
\centering
\includegraphics[width=9cm]{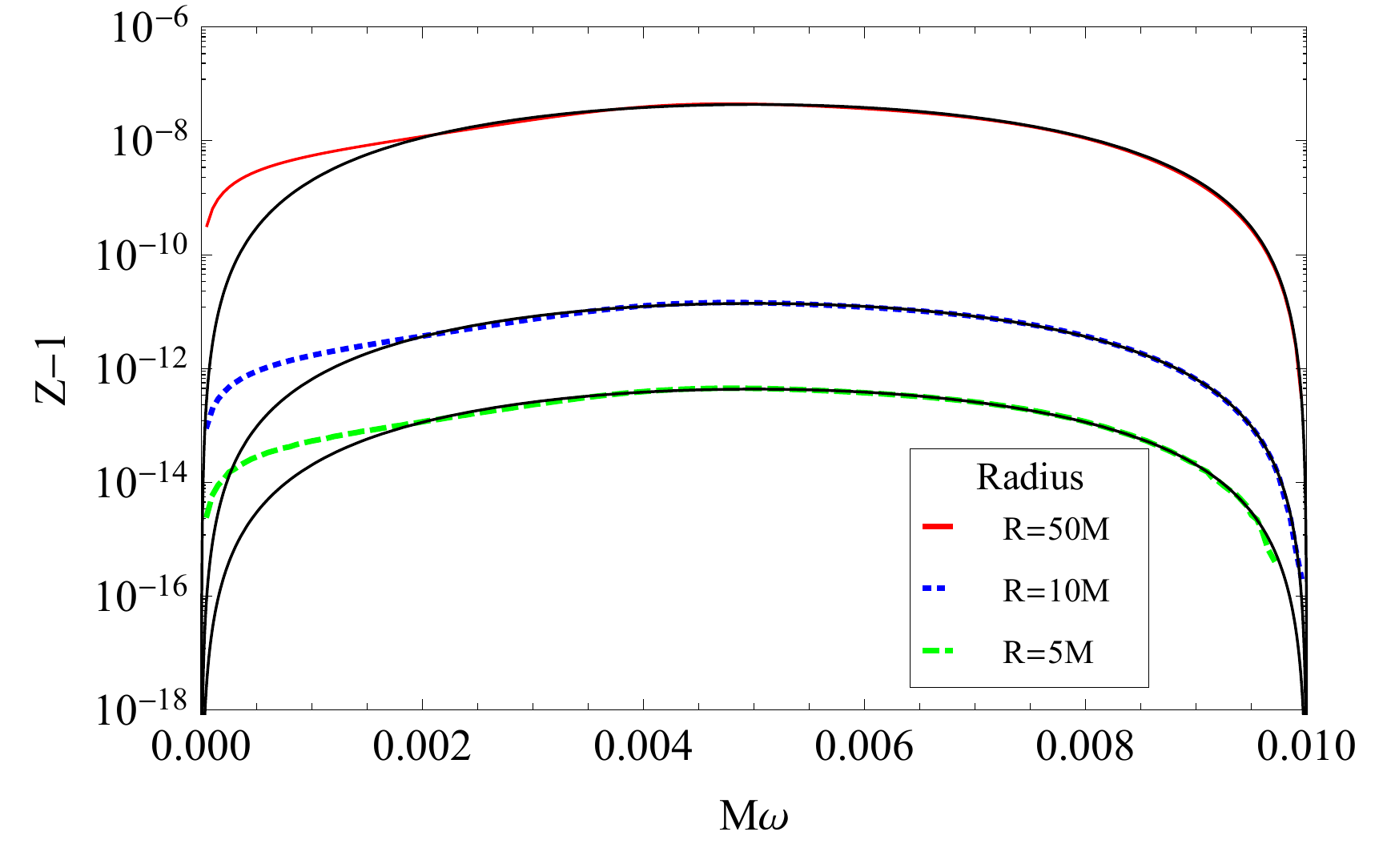}
\caption{Numerical and analytical amplification factors in the viscous analogue model.}
\label{ampvis}
\end{figure}

Thus, we will here briefly describe how our results change when dissipation inside the star is modeled through \eqref{visc}. We are here concerned only with rotational superradiance in flat spacetime, and we therefore take $\nabla^2$ to be the flat space-time Laplacian in spherical coordinates. The process of transforming to a co-rotating frame and the imposition of regularity at the center are identical to that of the model studied in chapter 3. The amplification factors were numerically obtained and are shown in figure \ref{ampvis}. 

Note that for low frequencies it is possible to map this analogue model into the previous toy model. Equation \eqref{visc} can be written as
\be
\left(1-\frac{4}{3}i\omega\nu\right)\box\phi+\frac{4}{3}i\omega\nu\partial_t^2\phi=0,
\ee
which in the limit $\omega<<1$ is the same as equation \eqref{modie} for $\alpha=\frac{4}{3}\nu\omega^2$. This map between $\alpha$ and $\nu$ allows us to obtain the non-relativistic analytical expression for the amplification factors:
\be
Z=1+\frac{16\nu R^2\left(\Omega-\omega\right)^3\left(\omega R\right)^{2l+1}}{3\left(2l+1\right)!!\left(2l+3\right)!!}.
\ee

It is worth noting that, if one were to try the naive black hole limit $R=2M$ with $\nu$ of order $M$, the above result would never recover the correct behavior for the amplification of waves by rotating black holes. The reason, it seems, is that dissipation in non-rotating black holes can be best described through \eqref{modie} rather than this viscous analogue. This is possibly due to the fact that the dispersion relation of equation \eqref{visc} in the eikonal limit leads to an energy dependent dissipative term \cite{liberati}.

\cleardoublepage

\end{document}